\documentclass[a4paper,pra,twocolumn]{revtex4-1}  

\usepackage{graphicx}  
\usepackage{dcolumn}   
\usepackage{bm}        
\usepackage{amssymb}   
\usepackage{hyperref}

\usepackage{soul,color}

\usepackage{caption} 
\usepackage{subcaption} 

\usepackage{amsfonts}

\usepackage{amsmath}
\usepackage{blkarray}
\usepackage{braket}
\usepackage{multirow}
\usepackage{mathtools}

\usepackage{tikz}
\usepackage[utf8]{inputenc}

\newcommand{\y}{\lambda}

\newcommand{\w}{\omega}

\newcommand{\dl}{\delta}

\newcommand{\sig}{\sigma}

\newcommand{\wrr}{\omega_R}

\newcommand{\stt}{\hat{\mathcal{S}}}
\newcommand{\ttt}{\hat{\mathcal{T}}}

\newcommand{\stket}[1]{{\mathcal{S}_{#1}}}
\newcommand{\ttket}[1]{{\mathcal{T}_{#1}}}

\newcommand{\nex}{\hat{\mathcal{N}}_{ex}}

\newcommand{\h}{\mathcal{\hat H}}

\newcommand{\hj}{\mathcal{\hat H}_j}

\newcommand{\hd}{\mathcal{\hat H}_{\Delta}}

\newcommand{\com}[1]{}

\newcommand{\eV}{~eV}

\newcommand{\ad}{\hat{a}^\dagger}
\newcommand{\an}{\hat{a}^{}}
\newcommand{\sd}{\hat{\sigma}^{+}_i}
\newcommand{\sn}{\hat{\sigma}^{-}_i}
\newcommand{\td}{\hat{\tau}^{+}_{i}}
\newcommand{\tn}{\hat{\tau}^{-}_{i}}

\newcommand{\fref}[1]{Fig.~(\ref{#1})}

\begin{document}


\title{Spin-orbit coupling in organic microcavities: Lower polariton splitting, triplet polaritons, and disorder-induced dark-states relaxation}

\author{M. Ahsan Zeb}
\affiliation{Department of Physics, Quaid-i-Azam University, Islamabad 45320, Pakistan}

\author{Shoaib Masood}
\affiliation{Department of Physics, Quaid-i-Azam University, Islamabad 45320, Pakistan}

\date{\today}

\begin{abstract}

Using an extended Tavis-Cummings model,
we study the effect of the spin-orbit coupling between the singlet and the triplet molecular excitons in organic microcavities in the strong coupling regime. 
The model is solved in the single excitation space
for polaritons, which contains the bright (permutation symmetric) 
singlet and triplet excitons,
as well as the dark bands consisting of the nonsymmetric
excitons of either type.
We find that the spin-orbit coupling splits the lower polariton into two branches, and also creates a \emph{triplet} polariton when the cavity mode is in resonance with the triplet excitons.
The optical absorption spectrum of the system that can reveal this splitting 
in experiments is presented and the effect of disorder in exciton energies and couplings is explored.
An important consequence of the disorder in the spin-orbit coupling ---
a weak coupling between the otherwise decoupled bright and dark sectors --- is explored and detailed calculations of the squared transition matrix elements between the dark bands and polaritons are presented
along with derivation of some approximate yet quite accurate analytical expressions.
This relaxation channel for the dark states 
contains an interference between two transition paths that,
 for a given polariton state,
suppresses the relaxation of one dark band and enhances it for the other. 

\end{abstract}

\maketitle

\section{Introduction}

The role of strong matter-light coupling~\cite{lidzey98} 
in modifying and determining various physical properties
has been extensively studied over the past two decades.
Some important examples include
the formation of polariton condensates~\cite{Kasprzak2006,Keeling2020:review} 
effect on superconducting phases~\cite{Mitrano2016,Schlawin2017},
charge and energy transport~\cite{Feist2015, Hagenmueller2017,Schafer2019,zeb2020},
and chemical reaction rates~\cite{Thomas2019,HerreraPRL16,Galego2016,galego17,MartinezACS2018,ebbesen2016hybrid,feist2017polaritonic,Ribeiro2018}.
Organic microcavities form an important class of systems that 
exhibit the strong matter-light coupling 
where the presence of the intramolecular vibrations and the triplet molecular excitons significantly enriches the prospects.
The vibronic coupling and polaronic effects in these systems 
 are very important and usually included in the models~\cite{nazir2016,zeb2020}.
However, the spin-orbit coupling between the singlet molecular exciton 
and the triplet molecular excitons 
in organic microcavities
has gained attention only very recently~\cite{polak20}.

The focus has been primarily on the energy landscape.
The triplet excitons in organic molecules lie below the singlet excitons 
(by the exchange splitting ${J\sim 0.5-1\eV}$)
and the singlet excitations relax to them (intersystem crossing, ISC~\cite{dias2013}).
However, thermally activated processes (reverse intersystem crossing, RISC~\cite{dias2013}) can transfer the excitations from the triplet states to the singlet states.
Compared to the singlet excitons that couple to the light and make polaritons, 
the
electronic pumping of organic microcavities creates three times as many triplets
and
it is desired to 
make them emissive or use them to increase the polariton population for condensation or lasing.
Thermally activated delayed fluorescence (TADF)~\cite{kaji2015,wong2017,Yin2020},
an exciting phenomenon that has attracted huge interest recently,
exploits the RISC.
RISC depends crucially on two quantities, (\emph{i}) the size of the energy barrier and (\emph{ii}) the transition matrix elements between the two types of excitations. 
The strength of the matter-light coupling in organic microcavities
that is experimentally achievable has increased over time and 
recently it was demonstrated to even bring the lower polariton 
(LP) below the triplet exciton state, the so called inverted regime~\cite{Eizner2019},
allowing even a barrier-free RISC~\cite{yu2021}.
The matrix elements can originate from the spin-orbit coupling and 
thus can be increased by using heavy metals 
that have a larger atomic spin-orbit coupling and hence render a greater singlet character to the triplet state and vice versa~\cite{baldo1998,adachi2001,kawamura2005,yersin2011}.
However, it does not matter how large the spin-orbit coupling is, the matrix elements still constitute a serious issue here. This is explained in the following.

For $N$ emitters coupled to a common cavity mode,
there is a single bright state that creates two polaritons and ${N-1}$ ``dark'' states that do not couple to the cavity at all~\cite{Ribeiro2018}.
In organic microcavities, 
this implies an overwhelmingly large density of states of such uncoupled singlet states that form an excitation reservoir~\cite{Ribeiro2018,Eizner2019,polak20}.
Even when the spin-orbit coupling is large, 
the polaritonic subsystem cannot fully benefit from the triplet excitations because, as we show later, they would predominantly populate the dark singlet reservoir. 
The only sizable matrix elements connecting the triplet sector to the polaritons exist for a single permutation symmetric triplet excitation. 
There are ${N-1}$ nonsymmetric triplets that all couple to the singlet reservoir only.
The fact that the high density of such dark singlet and triplet excitations dominates the dynamics and 
keeps the molecular excitations from the light emission has also been found recently~\cite{Eizner2019,MartinezJCP2019}.
These dark states 
relax to the emissive polaritonic state very slowly~\cite{Agranovich2002, Agranovich2003}. 
The intramolecular vibrations of organic molecules can act as a heat bath
and usually play an important role in   
the relaxation dynamics of excitations in organic microcavities~\cite{Litinskaya2004a,Michetti2009}.
In recent years, various theoretical~\cite{pino2018,groenhof2019,sommer2021} 
and experimental~\cite{Eizner2019,polak20,yu2021,mewes2020,wersal2019,xiang2019} 
works
explored the relaxation pathways and dynamics of polaritons and dark states
in these systems. 
It is interesting to explore new ways to relax the dark states to polaritons 
as it could help increase the polariton density for various applications.


\com{Since the cavity modes couple only to the singlet molecular excitons,
the spin-orbit coupling between the singlet and the triplet can indirectly couple the latter to the cavity mode.
}

\begin{figure}
\centering
\includegraphics[width=1\linewidth]{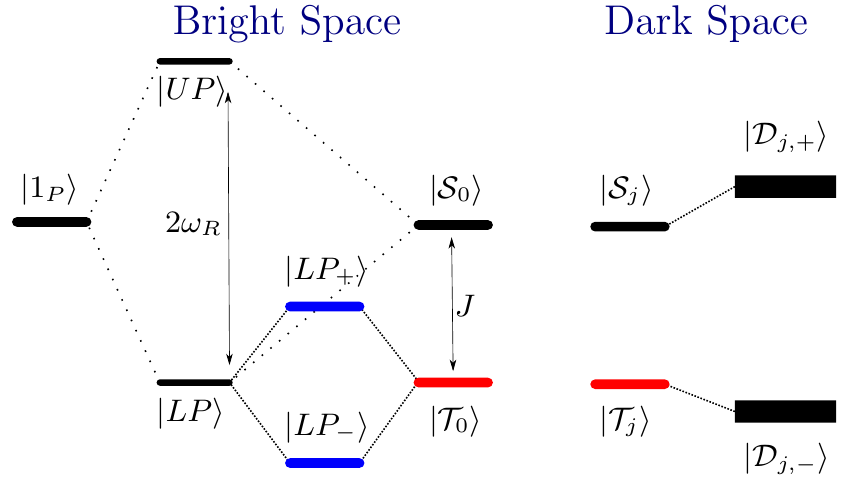}
\caption{A sketch of the energy level diagram showing
the cavity state $\ket{1_P}$ along with the singlet and triplet bright ($\ket{\mathcal{S}_0},\ket{\mathcal{T}_0}$) and dark ($\ket{\mathcal{S}_j},\ket{\mathcal{T}_j}$) molecular states.
The coupling between the singlet exciton and the cavity photon state
creates the lower and the upper polaritons, $\ket{LP}$ and UP.
Roughly speaking, the $\ket{LP}$  couples to the triplet molecular exciton state and produces 
LP$_{\pm}$.
The bright and the dark spaces that are decoupled for a clean system, as shown here,
get weakly coupled when a disorder in the spin-orbit coupling is present.
}
\label{fig:levels}
\end{figure}

In this work, we explore the 
effect of the spin-orbit coupling on the nature and dynamics of polaritons and the dark states in an organic microcavity
using an extended Tavis-Cummings (TC) model~\cite{TCM68}.
As shown in Fig.~\ref{fig:levels}, the $\ket{LP}$ originating from the coupling $\wrr$ between the (bright or symmetric) singlet exciton $\ket{\stket{0}}$ and the cavity mode can be made resonant to the triplet excitons as $\wrr$ approaches $J$. 
In such a case, the effect of the spin-orbit coupling between the $\ket{\stket{0}}$ component in $\ket{LP}$ and the symmetric triplet $\ket{\ttket{0}}$ becomes significant,
requiring to treat the two couplings 
(the light-matter coupling and the spin-orbit coupling) 
on an equal footing; i.e., the spin-orbit coupling should not be treated as a ``weak'' coupling.
It splits $\ket{LP}$ into $\ket{LP_\pm}$, both of which contain $\ket{\ttket{0}}$ along with $\ket{\stket{0}}$
but have no interaction with the dark states 
$\ket{\stket{j}}$ and $\ket{\ttket{j}}$
in a clean system.
It is known that an energetic disorder can 
couple $\ket{\stket{j}}$ to $\ket{\stket{0}}$
(leading to finite optical absorption by 
$\ket{\stket{j}}$~\cite{Houdre96,Cwik16}).
This is also true for
$\ket{\ttket{j}}$ and $\ket{\ttket{0}}$.
We will show in this article that a disorder in the spin-orbit coupling can weakly couple the bright and dark sectors of different character; i.e., it couples
$\ket{\stket{j}}$ to $\ket{\ttket{0}}$,
and 
$\ket{\ttket{j}}$ to $\ket{\stket{0}}$,
which induces transitions between them and hence can cause relaxation of the dark bands $\ket{\mathcal{D}_{j,\pm}}$ (that contain $\ket{\stket{j}}$ and $\ket{\ttket{j}}$) to, for instance, the lowest energy polaritons, $\ket{LP_-}$.
This could potentially be useful to increase the polariton population
for an enhanced incoherent light emission or, more importantly,
for creating a polariton condensate or laser.


In the following, we first consider a clean system in secs.~\ref{sec:model} and \ref{sec:results} to explore the effects of the spin-orbit coupling in organic microcavities.
We find that, similar to the TC model, we can decouple the system into bright and dark sectors, which make polaritons and dark bands in this case.
We find that the spin-orbit coupling splits the lower polariton $\ket{LP}$ into two branches $\ket{LP_\pm}$ and at a certain location in the dispersion, we can have \emph{triplet} polaritonic states.
The optical absorption spectrum of the system, which can show
$\ket{LP_\pm}$ in the experiments, is also presented.
In sec.~\ref{sec:disorder}, we consider
the robustness of the results of the clean model
against disorder 
both in energies and couplings (at typical disorder strengths)
by looking at its effects on the optical absorption spectrum.
Finally, in sec.~\ref{sec:relax},
we show that the disorder in the spin-orbit coupling 
weakly couples 
the bright and the dark sectors, which induces transitions between polaritons and the dark bands and thus opens a channel for
the dark states to relax into the lowest energy polaritons.
We perform detailed calculations of the corresponding transition matrix element
and present numerical results as well as accurate analytical expressions.

\com{The effects of disorder are considered in sec.~\ref{sec:disorder},
where we first 
confirm the robustness of the results of the clean model
against disorder both in energies and couplings (at typical disorder strengths)
by looking at its effects on the optical absorption spectrum.
}

\section{Model and calculations}
\label{sec:model}
Consider $N$ identical conjugated organic molecules placed inside a 
microcavity formed by, say, two plane mirrors with a separation 
that is of the order of half the wavelength that excites a singlet exciton
on the organic molecules.
This hybrid system 
couples the matter excitations to the cavity photons and can be excited optically or electronically.
We assume that only a single cavity mode couples to the molecules and
 the collective matter-light coupling $\wrr$ is larger than the losses, 
 the cavity leakage $\kappa$ and the nonradiative exciton decay 
 $\gamma$, that are inevitably present in real systems 
 (${\kappa\sim 0.07\eV}$~\cite{Cwik16}, ${\gamma\sim0.01\eV}$~\cite{Dimitrov16}), i.e., $\wrr\gg \kappa, \gamma$.
For simplicity,
we consider three electronic states of a molecule: $\ket{G},\ket{S} and \ket{T}$ --- the ground, singlet and triplet states. 
In this state space of a molecule $i$,
the creation and annihilation operators for the 
singlet ($\hat \sig_i^{\pm}$) and 
triplet ($\hat \tau_i^{\pm}$)
excitons can be defined as
\begin{align*}
\hat \sig_i^+ &= \begin{pmatrix}
0&0&0\\
1&0&0\\
0&0&0
\end{pmatrix},~~
\hat \tau_i^+ = \begin{pmatrix}
0&0&0\\
0&0&0\\
1&0&0
\end{pmatrix}, 
\\ 
\hat \sig_i^- &= (\hat \sig_i^+)^\dagger,~~
\hat \tau_i^- = (\hat \tau_i^+)^\dagger.
\end{align*}

\subsection{Hamiltonian in the rotating wave approximation}
\label{sec:}

The Hamiltonian of 
the system we consider 
can then be written as
\begin{multline}
\label{eq:hrwa}
\h
= 
\w_c \ad\an
   + 
  \sum_{i=1}^N \Big \{ \w_s  \hat \sig_i^+ \hat \sig_i^-
  +  \frac{\wrr}{\sqrt{N}}(\ad \sn + \sd \an)
   \\
+  \left[\w_t  \td\tn
  +
 \y (\sd\tn + \td\sn) \right] \Big\},
\end{multline}
where $\ad and \an$ are the creation and annihilation operators for the cavity photons, and
$\w_c$, $\w_s$ and $\w_t$ are the
energies of the cavity mode, the singlet exciton and the triplet exciton. 
$\wrr$ is the 
matter-light coupling (as mentioned before) whereas $\y$ is the spin-orbit coupling between the singlet and the triplet \emph{molecular} exciton states. 
If we consider all three triplets, $\ket{T}$ would be a symmetric superposition of them, while the two nonsymmetric superpositions would be decoupled.

Here, 
assuming that we are in the strong coupling
 regime~\cite{Skolnick1998},
the rotating wave approximation (RWA)~\cite{zubairy1997} has been made for the light-matter coupling that ignores 
the coupling between subspaces with different number of excitations $\nex$ 
(defined below), between states that 
have an energetic difference of $\sim 2\w_s$.
If $\wrr \ll \w_s$, these terms do not significantly affect the stationary
states of the system (due to a large energetic difference between the
 coupled states) and hence can be ignored.
Similar to the TC model~\cite{TCM68},
$\h$ commutes with the number of excitations, $\nex$,
where
\begin{gather}
\label{eq:nex}
\nex=\ad\an+\sum_{i} (\sd\sn + \td\tn),
\end{gather}
so we can diagonalize them simultaneously.
This allows diagonalization of $\h$ in a subspace with fixed $\nex$. 
Here we restrict ourselves to $\nex=1$ subspace.

The model above
ignores the intramolecular vibrations and their coupling to the electronic configurations~\cite{zeb2020}.
This approximation is well justified at large $N$ where 
polaron decoupling occurs~\cite{HerreraPRL16,zeb2020},
but, at $N\sim 1$, the vibronic coupling cannot be ignored 
and has to be included in the model~\cite{zeb2017,zeb2020}.
Since our main goal in this work is to study the effect of the spin-orbit coupling 
between the exciton states in organic microcavities,
we assume we are working with a large number of molecules, $N\gg 1$, where our model is valid. 
This amounts to ignoring the vibrational replicas 
of the polaritons and dark states that we discuss in this work.
We leave the polaronic effects at small numbers of molecules for future studies.

The model is versatile
as, for instance, 
the triplet molecular state in our model 
can also describe any other molecular state that interacts with the singlet coupled to the cavity mode but does not directly couple to the latter.
Specifically, it could be a 
vibrational state within the manifold of the electronic ground state, 
another singlet state with negligible dipole matrix element, 
or a charged molecular state~\cite{zeb2020}.
The interpretation of our results in such cases might also be interesting
but we will restrict our discussion within the intended scope in this paper.

In the following,
we first show that we can block-diagonalize $\h$ 
into polaritonic and dark sectors, where the latter can be analytically solved.
We then numerically diagonalize the polaritonic block
and find that the lower polariton splits into two branches, 
one of which becomes a purely triplet polariton in a small region of the dispersion, where exact expressions for all three polariton states can be obtained.
At the end, we present numerical results for the optical absorption spectrum.

\subsection{Block diagonalization of $\h$}
\label{sec:blockdiag}
The size of the ${\nex=1}$ 
subspace is ${2N+1}$
as there are $N$ states with a singlet exciton,
$N$ states with a triplet exciton and 
a single state with a cavity photon.
Noting that the spin-orbit term in the Hamiltonian (Eq.~\ref{eq:hrwa})
${\h_{soc} \equiv \y \sum_{i=1}^N(\sd\tn + \td\sn)}$
is diagonal in the molecular index,
we 
can apply any unitary transformation in the subspace of these $N$ singlet states
and keep the above structure of $\h_{soc}$ by 
applying the same transformation to $N$ triplet states as well.
That is, a transformation such as
\begin{gather}
\label{eq:stain}
\stt^{+}_{n}=\sum_{i=1}^N a_{i,n} \sd,\\
\ttt^{+}_{n}=\sum_{i=1}^N a_{i,n} \td,
\end{gather}
with the same coefficients $\{a_{i,n}\}$ (where $i,n\in[1,N]$),
will lead to
${\h_{soc} = \y \sum_{n=1}^N(\stt^{+}_{n}\ttt^{-}_{n} + H.c)}$,
where $H.c$ stands for Hermitian conjugate of the previous term.

What is the best choice for $\{a_{i,n}\}$?
We know that
using the permutation symmetric (bright) singlet state 
and the ${N-1}$ (dark) singlet states orthogonal to it
decouples the TC model
into corresponding bright and dark sectors
so using $\{a_{i,n}\}$ corresponding to this
bright-dark transformation
is guaranteed to 
simultaneously
blockdiagonalizes
the matter-light coupling and the spin-orbit coupling in
$\h$.
This brings a huge simplification as it
reduces the original ${(2N+1) \times (2N+1)}$ system 
to a simpler ${(2+1) \times (2+1)}$ bright polaritonic system along with 
${N-1}$ trivial ${2\times2}$ dark excitonic systems.
This is explicitly demonstrated in the following sections 
using a specific basis transformation.

\subsection{Bright and dark exciton states}
\label{sec:bright-dark}

We know that the symmetric (bright) and nonsymmetric (dark) superpositions of the singlet molecular exciton states block-diagonalize the TC model in the single excitation space reducing the problem to 
a two level polaritonic system along with $N-1$ dark states that are all decoupled from the cavity mode.
Following the discussion in the previous section,
we apply the same transformation to the triplet excitons as well.
Defining 
${\ket{GS}\equiv \ket{G_1,G_2,....,G_N}}$
to be the ground state of the molecules, 
the bright states,
${\ket{\stket{0}}\equiv\stt^+_0\ket{GS}}$ and ${\ket{\ttket{0}}\equiv\ttt^+_0\ket{GS}}$, 
are created by
\begin{eqnarray}
\label{eq:st0}
\stt^+_0 &\equiv& \frac{1}{\sqrt{N}}\sum_{j=1}^N\hat\sig_j^+,~
{\ttt}^+_0 \equiv \frac{1}{\sqrt{N}}\sum_{j=1}^N  
\hat\tau_j^+.
\end{eqnarray}

For the nonsymmetric or dark states, being a degenerate manifold, there is no unique representation and one can choose any complete set as basis states for the dark space.
We construct an orthonormal representation for the dark states
where different dark states tend to localize on different molecules, given by
${\ket{\stket{j}}\equiv\stt^+_j\ket{GS}}$,
and 
${\ket{\ttket{j}}\equiv\ttt^+_j\ket{GS}}$,
where ${j\in[1,N-1]}$
and the creation operators are,
\begin{eqnarray}
\label{eq:sj}
\stt^+_j &\equiv& \frac{1}{\sqrt{j(j+1)}} 
\left( \sum_{i=1}^j \hat\sig_i^+ - j \hat\sig_{j+1}^+  \right),\\
\label{eq:tj}
\ttt^+_j &\equiv& \frac{1}{\sqrt{j(j+1)}} 
\left( \sum_{i=1}^j \hat\tau_i^+ - j \hat\tau_{j+1}^+  \right).
\end{eqnarray}
As can be seen, 
the ${\ket{\stket{j}}}$ has a weight ${j/(j+1)}$ over a single molecule (labeled ${j+1}$)
and only a weight ${1/(j+1)}$ distributed equally over other $j$ molecules.
The reason we introduce this particular choice is that, 
 compared to more delocalized representations, e.g., Fourier transform, 
the ``nondiagonal'' matrix elements between the singlet and the triplet dark manifolds are suppressed 
when a disorder in the spin-orbit coupling is considered
(see sec.~\ref{sec:matelem}).

\subsection{Polaritons and dark bands}
\label{sec:}

We now show that,
as explained in sec.~\ref{sec:blockdiag},
the transformations in sec.~\ref{sec:bright-dark} above
block-diagonalize $\h$ in Eq.~\ref{eq:hrwa}.
In the single excitation space, 
we now have the following states,
$\{\ket{GS}\ket{1_P}, \ket{\stket{k}}\ket{0_P}, \ket{\ttket{k}}\ket{0_P}\}$,
where $\ket{0_P}~and ~\ket{1_P}$ are the cavity states with zero and one photons,
and $k=0,j$ with label $j$ is reserved for the dark states ($j\in[1,N-1]$).
We find that this particular choice of the basis states completely decouples $\h$ in
Eq.~\ref{eq:hrwa} into bright and dark parts,
\begin{gather}
\h = \h_B + \h_{D}.
\end{gather}

Here,
$\h_B$ is only a three state system in the bright space, i.e.,
$\{\ket{GS}\ket{1_P}, \ket{\stket{0}}\ket{0_P}, \ket{\ttket{0}}\ket{0_P}\}$,
given by
\begin{align}
\label{eq:hb}
\h_B =
\begin{pmatrix}
	\w_c 	 &  \wrr  &0\\  
	\wrr   &  \w_s &\y \\
	 0	&  \y 	     &\w_t
\end{pmatrix}
\equiv
\begin{pmatrix}
	\dl 	 &  \wrr  &0\\  
	\wrr   &  0 &\y \\
	 0	&  \y 	     & -J
\end{pmatrix},
\end{align}
where we 
define the detuning $\dl\equiv\w_c-\w_s$ and
exchange splitting $J\equiv\w_s-\w_t$,
and set $\w_s=0$ as the reference.
It is worth reminding that 
despite its simplicity, 
$\h_B$ above describes the \emph{collective} coupling of $N$ molecules to
the cavity mode including the spin-orbit coupling.
This is one of the main achievements of our transformations in Eqs.\ref{eq:st0}-\ref{eq:tj}.

We find that,
 $\h_{D}$ on the other hand is a sum of ${N-1}$ two-level-systems $\hj$
each in the dark space of 
$\{\ket{\stket{j}}\ket{0_P}, \ket{\ttket{j}}\ket{0_P}\}$.
The diagonalization of $\hj$ is trivial, it gives the same two energies for all ${j\in[1,N-1]}$ eigenstates due to identical coupling $\y$.
That is,
\begin{gather}
\h_D = \sum_{j=1}^{N-1} \hj,
\end{gather}
where 
\begin{align}
\label{eq:hd}
\hj =
\begin{pmatrix}
	\w_s 	 &  \y \\  
	\y   &  \w_t
\end{pmatrix}
&=
\begin{pmatrix}
	0 	 &  \y \\  
	\y   &  -J
\end{pmatrix},
\end{align}
with eigenstates and energies given by,
\begin{align}
\label{eq:dark2ls}
\ket{\mathcal{D}_{j,\pm}} &= A_{\pm} \ket{\stket{j}} \pm B_{\pm}\ket{\ttket{j}},\\
E_{j,\pm} &= \frac{1}{2}(-J \pm \sqrt{J^2+4\y^2}),\\
A_{+} &= \cos(\theta), A_{-} = \sin(\theta),
B_{\pm}=A_{\mp}, \\
2\theta &= \tan^{-1}(2\y/J),
\end{align}
where $\ket{0_P}$ is omitted for brevity.
We see that, in the limit $\y\ll J$, $\theta\sim0$ and 
$\ket{\mathcal{D}_{j,+}} \sim \ket{\stket{j}}$ while 
$\ket{\mathcal{D}_{j,-}} \sim -\ket{\ttket{j}}$.
Since, the energies $E_{j,\pm}$ are the same for all dark states in a given band,
these are in fact highly degenerate levels.
We will come back to the dark states later when we consider a disorder in the spin-orbit coupling, in sec.~\ref{sec:disorder}, where the energies $E_{j,\pm}$ make distributions of finite width.
In the following we explore
the ``bright'' sector represented by $\h_B$
that hosts more interesting polaritonic states.

First, we solve
$\h_B$ in Eq.~\ref{eq:hb} numerically 
for its eigenstates and energies to see how the spin-orbit coupling affects them. 
We find that
$\y\neq0$
splits the otherwise lower polariton $\ket{LP}$ into two branches, which 
we call $\ket{LP_\pm}$,
 while the upper polariton state is modified to a lesser extent.
The effect of the spin-orbit coupling is maximized when
the singlet and triplet states are brought closer in energy.
For a given $J$, this occurs when the
$\ket{LP}$ comes closer to the triplet,
i.e., at $\wrr\sim J$.
This will be shown in sec.~\ref{sec:split}.
At $\delta=-J$, the exact analytical solution is also possible, which shows a triplet polariton state, as will be discussed in sec.~\ref{sec:tp} along with the eigenstate structure in sec.~\ref{sec:struc}. 
For the experimentally relevant parameter 
regime ${\wrr,J \sim 0.5-1\eV}$, however,
we can approximately solve the model at any $\delta$ by
adiabatically eliminating the high energy state originating 
from the upper polariton $\ket{UP}$ to obtain an effective Two-level system
describing the low energy polaritonic branches of the spectrum.
We will show this in sec.~\ref{sec:approx}.
We will then present the numerical results for optical absorption in sec.~\ref{sec:abs}. 

\begin{figure}
\centering
\includegraphics[width=1\linewidth]{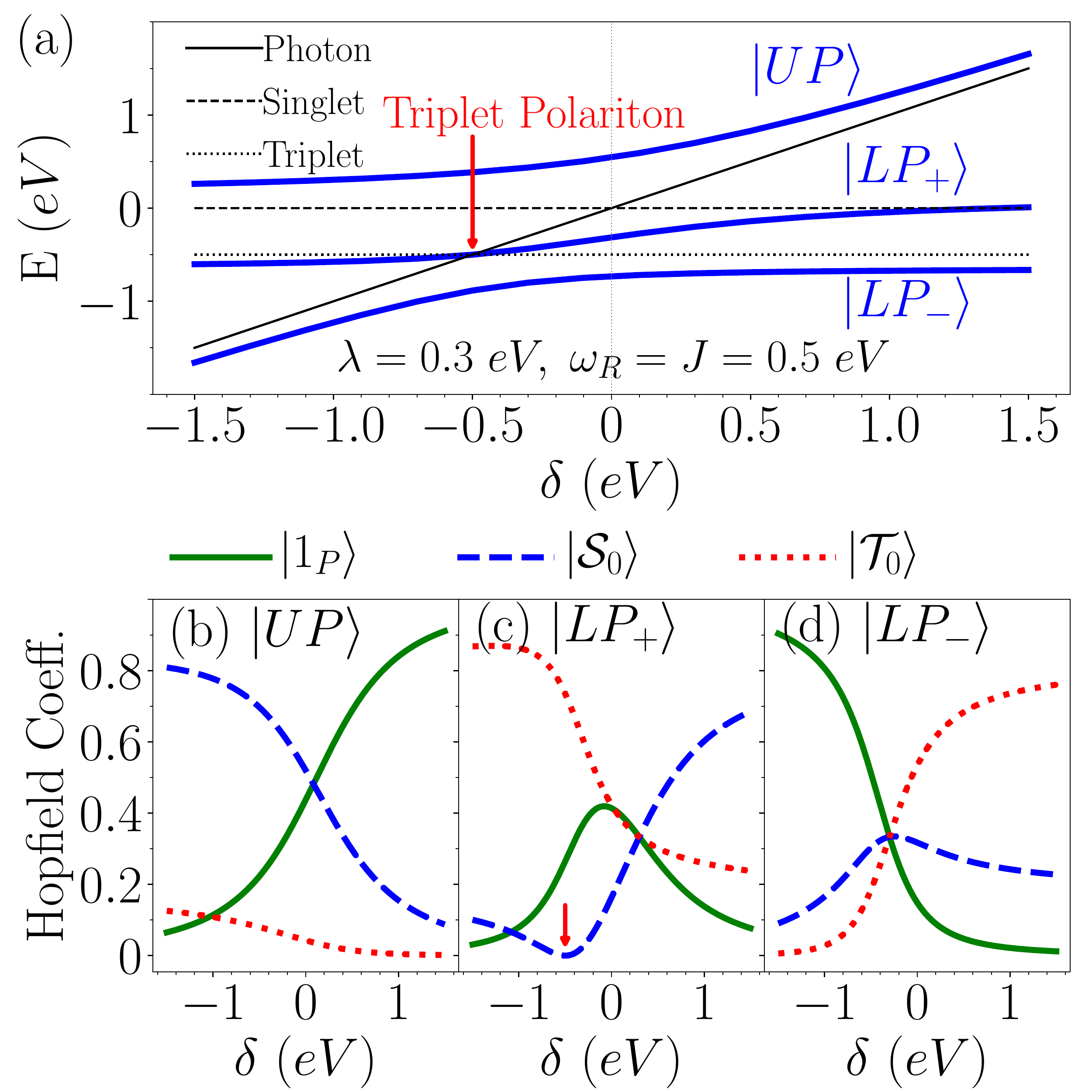}
\caption{The polaritonic states $\ket{UP}$ and $\ket{LP_\pm}$ at $\wrr=J=0.5\eV$ and $\y=0.3\eV$.
(a) The energy of $\ket{UP}$ and $\ket{LP_\pm}$ along with the 
bare energies of the three components --- the cavity photon and the molecular excitons. 
At $\delta=-J$, $\ket{LP_+}$ becomes a triplet polariton as indicated.
(b-d) The Hopfield coefficients, i.e., the weights of the component states.
The weight of the singlet in $\ket{LP_+}$ decreases to zero as $\delta\rightarrow-J$, the dip in the dashed line in (c) [indicated by a small arrow], is the region where triplet polaritons exist. }
\label{fig:wts}
\end{figure}

\section{Results and discussion}
\label{sec:results}

In this section, we will first present our results for the eigenstates and energies of the system and then the optical absorption spectrum.

\subsection{Stationary states and energies}
\label{sec:res-states}

In the following, we show how the spin-orbit coupling modifies the spectrum of the system. 

\subsubsection{Lower polariton splitting} 
\label{sec:split}

Figure~\ref{fig:wts}(a) shows the
spectrum of $\h_B$, i.e., the energy of the polaritonic states of the system
at ${\wrr=J=0.5\eV}$ and ${\y=0.3\eV}$, as a function of the  detuning $\delta$.
The energy of a bare singlet, triplet and photon is also plotted.
We call the highest energy state $\ket{UP}$
and the two lowest energy states 
$\ket{LP_\pm}$, due to their relation to the upper polariton and lower polariton of the TC model (i.e., the $\y=0$ case).
In \fref{fig:wts}(a), two anticrossings can be observed: one between $\ket{UP}$ and $\ket{LP_+}$
near zero, the energy of the singlet state,
and one between $\ket{LP_\pm}$ near the energy of the triplet state, $-J=-0.5\eV$.
$\ket{LP_\pm}$ can be thought of as arising from the splitting of the 
$\ket{LP}$ due to its interaction with the triplet state, hence the names.

In the following, we focus at  
$\delta=-J$, the point where the cavity dispersion crosses the triplet state.
There, the $\ket{LP_+}$ state completely loses its singlet exciton component,
becoming a triplet-only polariton, as marked in Figs.~\ref{fig:wts}(a,c).

\subsubsection{Triplet Polaritons}
\label{sec:tp}

$\ket{LP_+}$ becomes a purely \emph{triplet polariton}
when the cavity mode is in resonance with the triplet state,
 i.e., at $\w_c=\w_t$ or $\dl=-J$.
At this point, 
$\h_B$ can be solved analytically by yet another bright-dark transformation, 
involving weighted symmetric and nonsymmetric superpositions
of the photon and the triplet state,
${(\wrr \ket{1_P} + \y \ket{\ttket{0}})}$ and ${(\y \ket{1_P} - \wrr \ket{\ttket{0}})}$,
where the symmetric state couples to the singlet with an enhanced
strength $\sqrt{\wrr^2+\y^2}$
but the nonsymmetric state is completely decoupled from it. 
The symmetric superposition
 forms $\ket{LP_-}$ and $\ket{UP}$ eigenstates, 
 given by (unnormalized),
\begin{align}
\label{eq:tpsym}
\ket{LP_-} &= \wrr \ket{1_P} + \y \ket{\ttket{0}} + E_{UP}\ket{\stket{0}},\\
\ket{UP} &= \wrr \ket{1_P} + \y \ket{\ttket{0}} + E_{LP_-}\ket{\stket{0}},
\end{align}
with energies
\begin{align}
\label{eq:tpsyme}
E_{LP_-} &= 
\frac{1}{2}\left(-J -\sqrt{J^2+4 \left(\lambda ^2+\wrr^2\right)}\right),\\
E_{UP} &= 
\frac{1}{2}\left(-J+\sqrt{J^2+4 \left(\lambda ^2+\wrr^2\right)}\right),
\end{align}
while the nonsymmetric state 
is an eigenstate of $\h_B$, given by
\begin{align}
\label{eq:trippol}
\ket{LP_+}_T &= 
\frac{\y \ket{1_P} - \wrr \ket{\ttket{0}}} 
{\sqrt{\y^2+\wrr^2}}.
\end{align}
It is worth noting that even though $\ket{LP_+}_T$ forms at the same energy ${E=\w_c=\w_t}$, 
it is not a trivial superposition of its components, rather it is a genuinely new nondegenerate state orthogonal to the other eigenstates of the system.
Furthermore, from the exact numerical results at ${\delta\neq-J}$, we find that in a 
small region around this resonance,
the weight of the singlet exciton vanishes up to first order.
This means that the $\ket{LP_+}$ state is approximately a triplet polariton
 in a small but finite region. This will be explained in sec.~\ref{sec:struc}.
From Eq.~\ref{eq:trippol}, we can see that the maximal mixing occurs at $\y=\wrr$,
where the photon $\ket{1_P}$ and the bright triplet $\ket{\ttket{0}}$ have equal weights.
We can see from Eq.~\ref{eq:trippol} that
$\ket{LP_+}_T$ becomes more triplet like at $\y<\wrr$ and more 
photon like at $\y>\wrr$.

\subsubsection{Eigenstate structure}
\label{sec:struc}

The three polaritonic eigenstates 
of $\h_B$ in general contain all three components:
the photon, the bright singlet and the bright triplet excitons.
For a given eigenstate, we call the weights of these components its Hopfield coefficients.
Figures~\ref{fig:wts}(b-d)
shows the Hopfield coefficients for the three eigenstates
for the same parameters as in \fref{fig:wts}(a).
$\ket{UP}$ is far away from the triplet state
so 
there is a negligible triplet component in this branch, 
where as its singlet and photon components 
are only slightly affected by a finite spin-orbit coupling $\y$.
Close to resonance ($\dl=0$), 
all states have 
a nonzero photon component.
$\ket{LP_\pm}$
both contain a sizable triplet component.
$\ket{LP_+}$ contains the photon close to the resonance only 
and it approaches to excitons away from it.
The singlet component in
$\ket{LP_-}$ decreases away from the resonance but it still stays relatively large. On the other hand this state 
approaches the photon or the triplet away from the resonance.

As indicated by a small arrow in Fig.~\ref{fig:wts}(c),
the singlet weight in $\ket{LP_+}$ has \emph{a minimum} that
touches zero at $\delta=-J$ making $\ket{LP_+}$ a pure triplet polariton.
Since around a minimum of any function, there is no first order change and the leading term is second order,
there is a small region $\delta\w$ around this point where
the singlet weight in $\ket{LP_+}$ vanishes up to first order in $\delta\w/2$
and so the triplet polariton picture is valid to this order of approximation.

From the structure of $\ket{LP_-}$,
one can naively think that 
it might help the triplet to singlet transition
that is highly desirable for optoelectronic devices.
However, as Ref.~\cite{MartinezJCP2019} shows, 
the density of the dark states $\ket{\stket{j}}$ and $\ket{\ttket{j}}$,
or their dark bands that are eigenstates of $\h_D$, is much larger and determines the reverse intersystem crossing rates.
In other words,
it is only the permutation symmetric superposition of the singlet and triplet, $\ket{\stket{0}}$ and $\ket{\ttket{0}}$,
that make up the polaritonic states.
However, as we show in sec.~\ref{sec:disorder}, a disorder in 
the spin-orbit coupling can open up a relaxation channel for the dark bands.

\begin{figure}
\centering
   \includegraphics[width=1\linewidth]{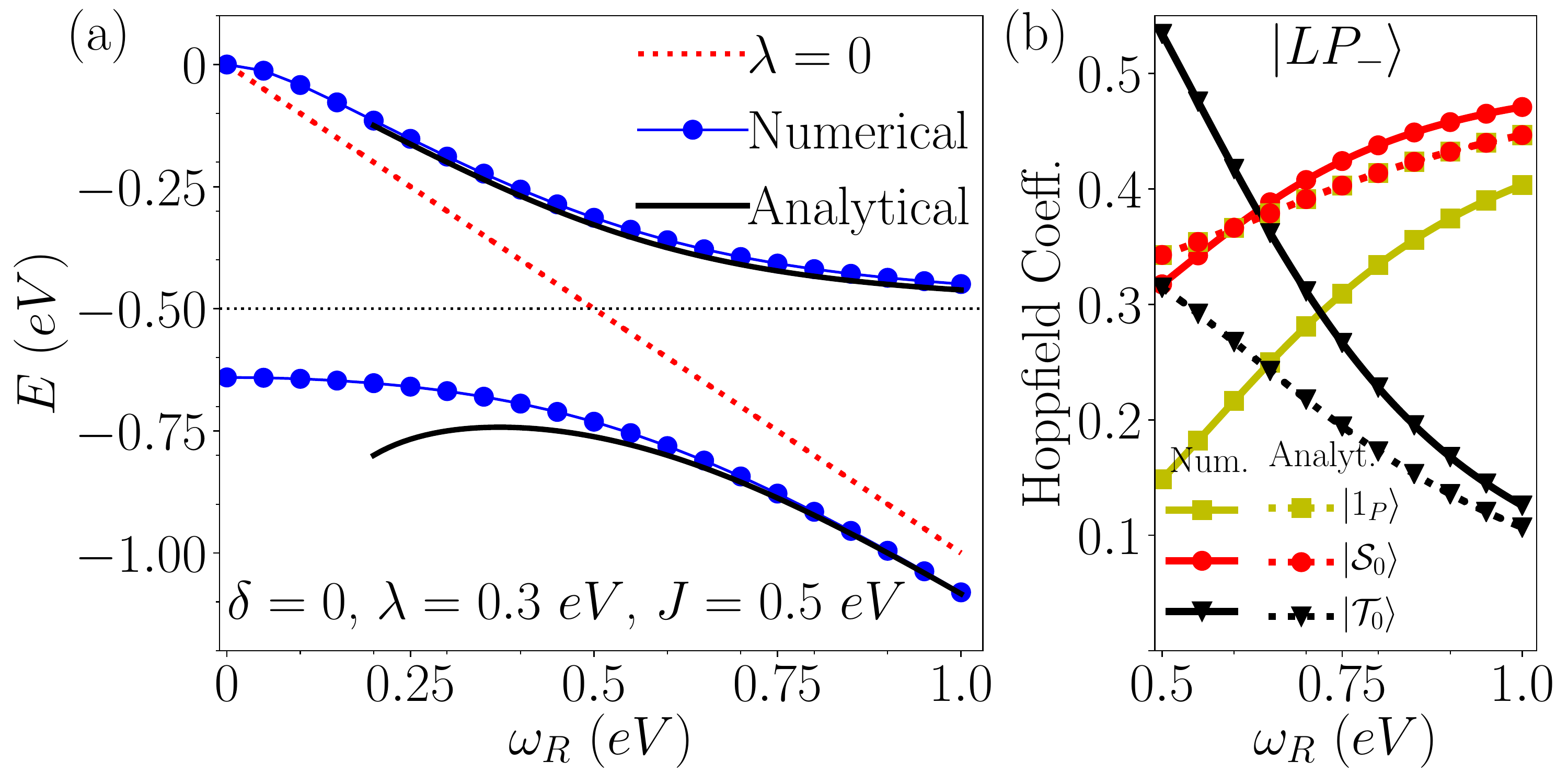}
\caption{Comparison of approximate analytical and numerical results for the low energy eigenstates at $\delta=0$, $\y=0.3\eV$ and $J=0.5\eV$.
(a)
The energies of $\ket{LP_\pm}$ as a function of $\wrr$.
The triplet energy $-J$ as well as the $\ket{LP}$  energy at $\y=0$, i.e., $-\wrr$, are also shown as dotted lines.
(b) The Hopfield coefficients of the $\ket{LP_-}$ at $\wrr\geq J$. 
The analytical results give the same weight to the photon and the bright singlet states (at $\delta=0$) because of the way they are constructed using the adiabatic elimination.}
\label{fig:res}
\end{figure}

\subsubsection{Approximate expressions for $\ket{LP_{\pm}}$}
\label{sec:approx}

At $\y=0$, $\h_B$ can be solved analytically, giving the two singlet exciton polaritons of the TC model
and the decoupled triplet exciton state.
A unitary transformation to these basis states
can indicate the origin of the lower polariton
splitting 
at $\y\neq0$
and help us obtain an approximate analytical
result for the two low energy branches, $\ket{LP_\pm}$.
Let us take the resonant case, $\dl=0$,
to illustrate this point.
In this basis,
$\h_B$ can be written as
\begin{equation}
\label{hb-tcpol}
\h_B=
\begin{pmatrix}
	+\wrr & 0& \frac{\y}{\sqrt{2}}\\  
	0&-\wrr & \frac{\y}{\sqrt{2}} \\
	\frac{\y}{\sqrt{2}}&  \frac{\y}{\sqrt{2}}&-J	
\end{pmatrix},
\end{equation}
which, on adiabatic elimination~\cite{Brion-2007} of 
the $\ket{UP}$ (the high energy state
at $+\wrr$), reduces to
\begin{equation}
\label{eq:hbred}
\h_B=
\begin{pmatrix}
	-\wrr & \frac{\y}{\sqrt{2}} \\
	\frac{\y}{\sqrt{2}}&-J-\frac{\y^2}{2\wrr}	
\end{pmatrix}.
\end{equation}
This reduced model of the two lower energy states
is a good approximation in the parameter regime, $\wrr\gtrsim J$,
as we will see shortly.

At ${\delta\sim0}$ and ${\wrr\sim J}$, the parameter regime we considered here,
diagonalizing the matter-light part first to form the polaritonic states has an advantage over 
diagonalizing the matter-only spin-orbit term first.
In the former, one state gets closer to the left over state (triplet) and 
one gets away from it, while in the latter case, both mixed matter states
 get away from the left over state (photon state).
Thus to obtain the two lowest energy states $LP_{\pm}$, 
adiabatic elimination is more accurate in the former case 
where we eliminate the state that is away from the other two.
Had we taken the eigenstates at $\wrr=0$, i.e., 
had we diagonalized the spin-orbit term first,
we would have obtained two exciton states with one above the singlet and the other below the triplet. 
At $\delta=0$, it would have
allowed us to eliminate the \emph{lower} state 
due to an extra energetic difference of $J$ 
 to obtain 
approximations for the two high energy eigenstates LP$_{+}$ and UP, 
but not the lowest energy state $\ket{LP_-}$ that we are interested in.
Nonetheless, different routes can be necessary for different domains of $\delta$, e.g., diagonalizing the exciton states first can work better at $\delta<-J$ where the photon state is closer to the lower excitonic state permitting the elimination of the high energy state.

Coming back to Eq.~\ref{eq:hbred},
the eigenstates (unnormalized) of the reduced model and their energies are given by,
\begin{align}
\label{eq:analv}
\ket{LP_\pm} &=
\alpha_{LP_{\pm}} \ket{\stket{0}} +
\beta_{LP_{\pm}}\ket{\ttket{0}} +
\gamma_{LP_{\pm}} \ket{1_P},\\ \nonumber
\alpha_{LP_{\pm}} &=  
\frac{1}{\sqrt{2}}\left(E_{\pm}+J+\y^2/2\wrr\right),\\ \nonumber
\beta_{LP_{\pm}} &= \y/\sqrt{2},\\ \nonumber
\gamma_{LP_{\pm}} &= - \alpha_{LP_{\pm}} ,\\
\label{eq:anal}
E_{\pm} &= {-\Delta \pm \sqrt{\Delta^2-J \wrr}},\\
\nonumber
\Delta &=(\wrr +J)/2 +{\y}^2/4\wrr.
\end{align}
Figure~\ref{fig:res}(a) compares Eq.~\ref{eq:anal} and the exact numerical low energy spectrum at $\dl=0$, $\y= 0.3 \eV$, and $J=0.5 \eV$. 
The red dotted line shows the energy of the 
$\ket{LP}$ when there is no spin-orbit coupling ($\y=0$).
The analytical results agree reasonably well to the exact numerical results at $\wrr\gtrsim J$.
The approximate eigenstates in Eq.~\ref{eq:analv} are compared to the exact results in Fig.~\ref{fig:res}(b). 
The approximate results force the singlet and the photon weights to be equal
 (squares partially hidden behind the circles in the upper dotted line), 
 which becomes closer to the exact results at $\wrr\sim1 eV$.
Compared to the energies in Fig.~\ref{fig:res}(a), 
the approximate state 
requires a relatively larger $\wrr$ ($\wrr>1 eV$, not shown) 
for a similar match to the exact results.

We have seen how $\y>0$ affects the stationary states of the system.
Let us consider something that is routinely measured in the experiments, the optical absorption spectrum.

\begin{figure}
\centering
  \includegraphics[width=1\linewidth]{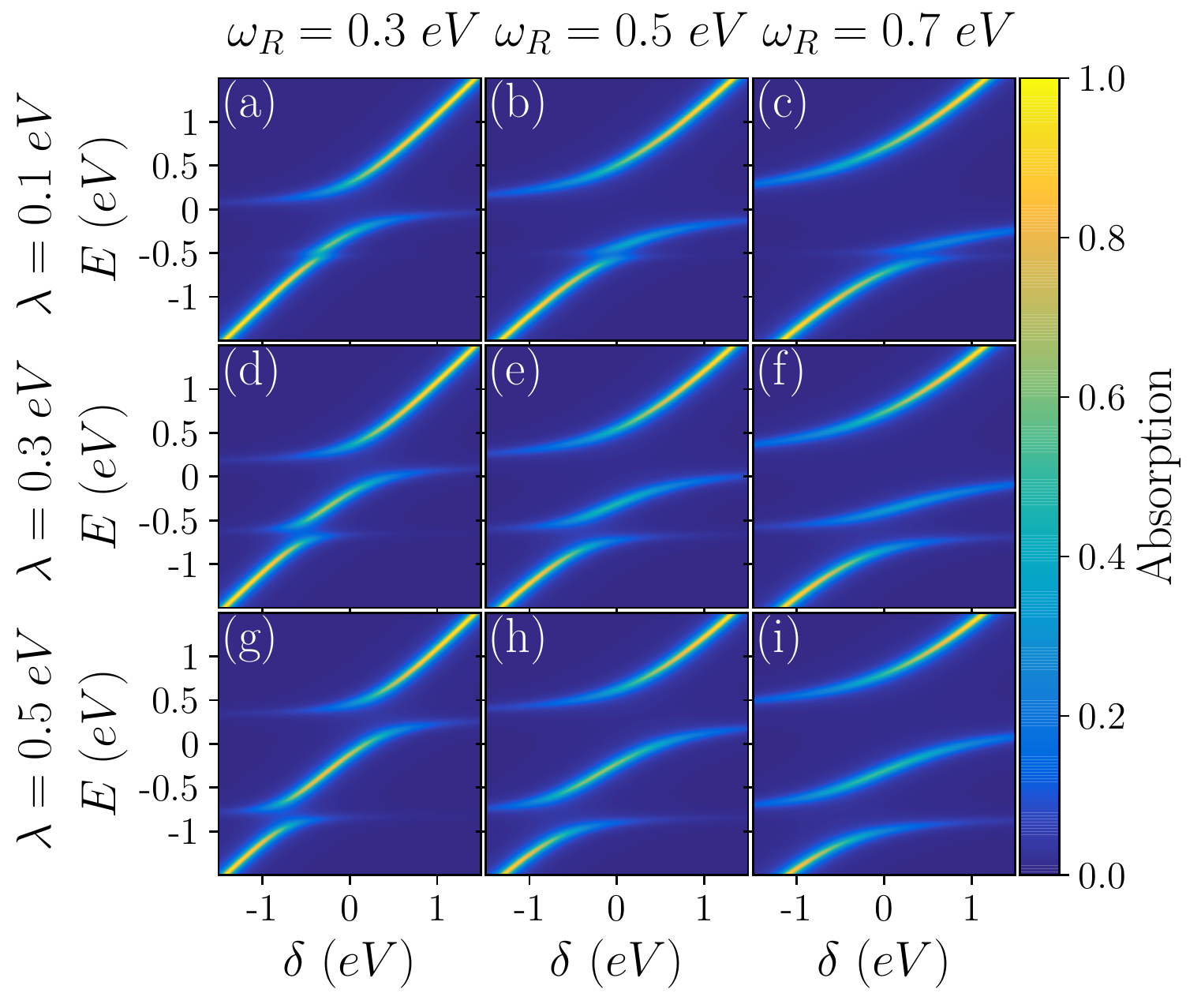}
\caption{Absorption spectrum at $J=0.5 \eV$. 
The matter-light coupling $\wrr$ varies along the row where as the spin-orbit coupling $\y$ changes along the column.
At fixed $\wrr$ (in a column), the lower polariton splits into two branches $\ket{LP_\pm}$ as we move to lower rows to increase $\y$.
The effect becomes more prominent as we increase $\wrr$ and bring the $\ket{LP}$  energy closer to that of the triplet exciton.
}
\label{fig:abs}
\end{figure}

\subsection{Absorption Spectrum}
\label{sec:abs}

In experiments, measuring the optical absorption 
by the organic microcavities
is a standard method to investigate strong matter-light coupling and polaritons. 
In this section, we present the calculated absorption spectrum of our system
for a selected set of $\wrr$ and $\y$ values.
Figure~\ref{fig:abs} shows the absorption spectrum versus detuning $\dl$
at $J=0.5 \eV$.
The value of $\wrr$ varies from $0.3 \eV$ to $0.7 \eV$ along the row from the left to the right, 
and $\y$ varies from $0.1 $  to $0.5 \eV$
from the top to the bottom along the column.
Starting from the top left,
\fref{fig:abs}(a),
and moving to the right, 
we see the very faint effect of the triplet
and the absorption appears almost similar to
that of the singlet polaritons of the TC model.
However,
as we increase $\y$ to $0.3 \eV$, moving to the second row,
the second anticrossing now becomes prominent
and we clearly see the splitting of the 
LP into two branches.
The splitting increases as we increase $\wrr$ because 
it decreases the effective detuning between the singlet polariton and the triplet state.
Increasing $\y$ further to $0.5 \eV$ in the bottom row,
this splitting increases accordingly
and in \fref{fig:abs}(i), the two branches of the 
$\ket{LP}$, $\ket{LP_\pm}$ even become deceptive --- they could 
be found in the experiments but look similar to the usual singlet polaritons
(except that they both become flat at large positive $\delta$).
Looking at columns, from the top to the bottom,
i.e., at fixed $\wrr$, 
we observe the 
$\ket{LP}$ splitting into $\ket{LP_\pm}$
as $\y$ increases.

Since energies and couplings in organic materials are usually disordered, 
the robustness of the picture presented so far against disorder needs to be checked. This will be done in the following section.


\section{Effects of disorder}
\label{sec:disorder}

In this section, we discuss the effects of the disorder that is inevitably present in organic materials.
The energies $\w_s,\w_t$ of the singlet and the triplet exciton states 
in organic materials contain disorder and form a distribution 
of finite width around a mean value.
The strength of this energetic disorder can be characterized as the width
(standard deviation) $\Delta_e$ of a Gaussian distribution, 
and its typical value is
$\Delta_e\sim0.1 eV$~\cite{bassler81, bassler82}.
A similar disorder in the spin-orbit coupling $\y$ is bound to happen 
due to differences in the individual molecular exciton states,
while the randomness in the orientations of molecules 
and their position in the cavity relative to the antinode of the
 electromagnetic field in the cavity leads to a distribution in 
 the light-matter coupling $g$ for individual molecules.

\subsection{Model including disorder}
\label{sec:modeld}

Considering such disorder, 
the energies 
of the singlet and the triplet molecular states 
are no longer
fixed at $\w_s,\w_t$ but are \emph{randomly} taken from a Gaussian 
distribution of width $\Delta_e$ centered around these.
Similarly, the spin-orbit coupling and the coupling of the individual molecules to the cavity mode
are drawn randomly from Gaussian distributions of width 
$\Delta_\y$ and $\Delta_g$ centered around $\y$ and $g=\wrr/\sqrt{N}$, respectively. 
Here, we impose a constraint 
$\sqrt{\sum_{i=1}^N {g_i^2}} = \wrr$ to keep the collective coupling to a fixed value because it determines the polariton energies in the absence of disorder and can serve as a reference.
This also allows us to use $\Delta_{\wrr}\equiv\sqrt{N}\Delta_g$ that avoids the trivial dependence on $N$.
The combined notation we use for the coupling disorder width is $\Delta_c$ with $c=\y,\wrr$.

The Hamiltonian $\h$ in Eq.~\ref{eq:hrwa} modifies to $\hd$, given by
\begin{multline}
\label{eq:hd}
\hd
= 
\w_c \ad\an
   + 
  \sum_{i=1}^N \Big \{ \w_{s,i}  \hat \sig_i^+ \hat \sig_i^-
  +  g_i(\ad \sn + \sd \an)
   \\
+  \left[\w_{t,i}  \td\tn
  +
 \y_i (\sd\tn + \td\sn) \right] \Big\},
 \end{multline}
where $\w_{s,i},\w_{t,i},g_i, and \y_i$ are energies and couplings for the $i$th molecule.
Due to the presence of disorder in the diagonal terms in $\hd$ in Eq.~\ref{eq:hd},
the delocalized symmetric singlet and triplet states can no longer reduce
the problem size and we have to work in the full Hilbert subspace for a single excitation. That is,
we cannot separate the bright and the dark excitonic spaces.
We numerically diagonalize $\hd$ and calculate the absorption spectrum to see how robust our previous findings are against the disorder. The results are presented in the following section.

\subsection{Robustness of $LP_{\pm}$}
\label{sec:disabs}

\begin{figure}
\centering
  \includegraphics[width=1\linewidth]{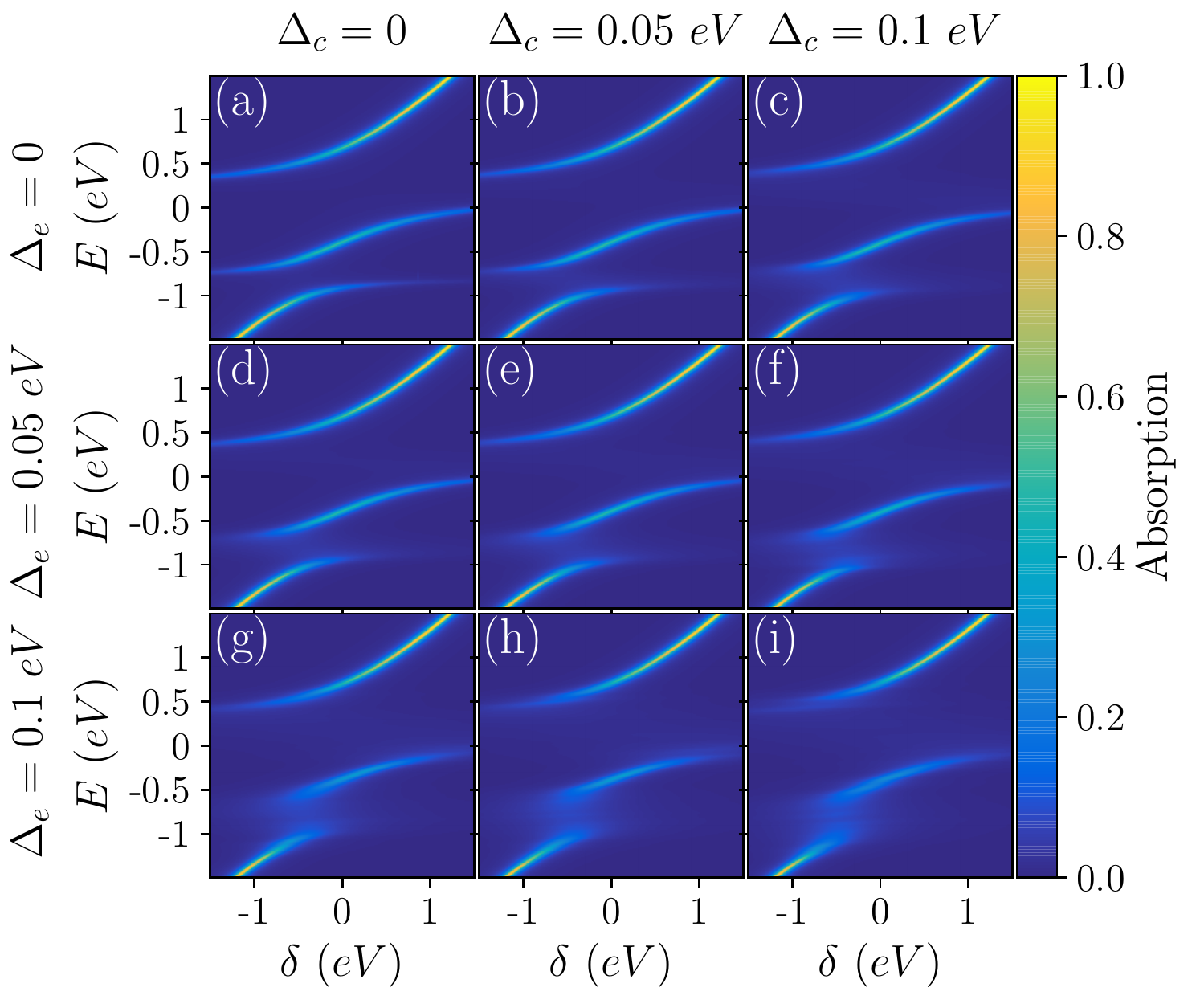}
\caption{Effects of disorder in exciton energies and couplings on the optical absorption spectrum.
Energies and couplings are taken randomly from gaussian distributions of widths $\Delta_e$ and $\Delta_c$
with mean values given by $\w_s=0, \w_t=-J=-0.5\eV$, and $ \w_R=0.5\eV, \y=0.3\eV$.
$\Delta_e$ varies along the column whereas $\Delta_c$ varies along the rows,
as labelled.
Panel $(a)$ is the same as Fig.~\ref{fig:abs}$(e)$.
The results are obtained at $N=100$ and are 
averaged over $400$ independent random samples for the energies and couplings
to reduce the noise due to a relatively small $N$.
We see that the energetic disorder is relatively more detrimental but the spectrum
keeps its main features intact even at the largest (and quite realistic) disorder strengths shown [panel $(i)$].
}
\label{fig:disorder}
\end{figure}

\emph{Absorption spectrum with disorder}: Figure~\ref{fig:disorder} shows the absorption spectrum at three values of disorder in energies and couplings, ${\Delta_{e/c}=0,0.05,0.1~ eV}$,
calculated at ${N=100}$ and averaged over $400$ random realizations of the energies and couplings to remove the noise at this relatively small $N$. 
$\Delta_{e}$ increase from the top to the bottom while 
$\Delta_{c}$ increase from the left to the right.
The other parameters are the same as in Fig.~\ref{fig:abs}(e),
so that at ${\Delta_{e}=\Delta_{c}=0}$ [Fig.~\ref{fig:disorder}(a)],
the results match those of the clean system.
First, we note that, compared to the disorder in the couplings $\Delta_c$,
 the energetic disorder $\Delta_e$ spreads the photon spectral weight 
 to a greater extent. This can be seen by comparing the $i$th column 
 to the $i$th row in Fig.~\ref{fig:disorder} [e.g., for the first row and column, Figs. (d) and (g) 
 are affected more than Figs (b) and (c)].
The $\ket{LP_-}$ branch that is very faint at $\delta \gtrsim 0.3 eV$
even without disorder, quickly disappears in this and neighboring 
region when either $\Delta_{e}$ or $\Delta_{c}$ are $0.1eV$, 
see the last row and columns in Fig.~\ref{fig:disorder}.
However, we see that the two anticrossings 
are quite intact even when both the disorder types
 reach their maximum strength, ${\Delta_{e/c}=0.1eV}$, 
 where they are relatively diffused but still recognizable.
Thus, the overall picture of lower polariton splitting should still be identifiable in the experiments involving typical disorder strengths shown here.

In the following, we consider the consequence of the disorder in the spin-orbit coupling on the dynamics of the dark bands and polaritons. We show that it opens up a relaxation channel for the dark states.

\section{Relaxation dynamics of the dark states}

\label{sec:relax}

We know the basic difference between the generic weak and the strong coupling cases.
In the former case, the stationary states are not much different from the original states, they acquire only a small component from each other
that causes transitions between the original states usually relaxing the system towards the lower energy.
(In the latter case, the stationary states are superpositions of the original states shifted in energies with sizable components from multiple original  states.)
Inter-system-crossing and reverse intersystem crossing
are examples of the weak coupling in organic materials
that is caused by a spin-orbit coupling $\y$ that is much smaller than 
the energetic difference $J$ between the singlet and the triplet exciton states and hence is assumed to couple them only weakly.

In the clean system we considered in 
{Secs.~\ref{sec:model} and \ref{sec:results}},
the bright and the dark states are energetically separated.
The bright states along with the cavity states make the polaritons 
while the dark states in the singlet and the triplet sector 
are completely decoupled from the bright sector.
They make the dark bands or levels due to their intersector coupling.

\com{
In the absence of the spin-orbit coupling,
we know from the Tavis-Cummings model that the energetic disorder $\Delta_e$ introduces coupling between the bright and the dark singlet states and induces photon spectral weight at the bare singlet exciton energy that is visible in the optical spectra of organic microcavities.
It is interesting to see that 
 in the disorder in the spin-orbit coupling 
can couple the bright singlet state to the dark triplet sector, as we show below.
}

We would like to explore how a disorder in the spin-orbit coupling affects this picture.
We find that it
can couple the bright singlet states to the dark triplet states
and vice versa.
This weak coupling
will induce transitions between the dark bands and polaritons
where the former
contain the dark components and the latter the bright components.
So we have a relaxation channel that can transfer the excitations 
from the dark bands to the polaritons.

To compute the transition rate between two states, 
we need to consider two quantities, 
the squared transition matrix element between them
and the energy difference between them.
If we are to compute the transition rate from a set of degenerate states to a given state,
we can combine the squared transition matrix element for all states in the set.
Since the dark bands should not be too wide,
we can apply this idea here and
combine the squared transition matrix elements to get a single number that can be used in the Fermi golden rule, or a more sophisticated formalism for an open quantum system involving the bath spectral density~\cite{Pino15,pino2018}, 
to calculate the total transition rate.

It can be easily verified that
for a degenerate manifold where multiple representations of the states are possible,
the transition matrix elements squared for all states always sum up to a unique value and it is independent of the choice of the representation.
We will continue using the same representation for the dark states that we introduced in
sec.~\ref{sec:bright-dark}
to compute this quantity.

\subsection{Matrix elements}
\label{sec:matelem}

Let us calculate
the matrix elements of the Hamiltonian $\hd$ in Eq.~\ref{eq:hd} containing a disorder in the spin-orbit couplings only, i.e., $\Delta_{\y}>0$, but ${\Delta_{e}=0=\Delta_{\wrr}}$.

\begin{align}
\nonumber
\mathcal{H}_{\stket{0},\ttket{0}} &=
\braket{\stket{0}|\hd |\ttket{0}},\\
&=
\frac{1}{N}\sum_{i=1}^{N} \y_i=\braket{\y}=\y,
\end{align}
where $\braket{\y}=\y$ is the mean value of the spin-orbit coupling.
Defining,
\begin{align}
\Lambda_j&\equiv\left( \sum_{i=1}^{j} \y_i - j \y_{j+1} \right),
\end{align}
we can similarly write other nonzero matrix elements,
\begin{align}
\label{eq:m0j}
\mathcal{H}_{\stket{0},\ttket{j}} &=
\frac{1}{\sqrt{N j(j+1)}}\Lambda_j
=\mathcal{H}_{\ttket{0},\stket{j}},\\
\label{eq:mjk1}
\mathcal{H}_{\stket{j},\ttket{j'>j}} &=
\frac{1}{\sqrt{j(j+1)j'(j'+1)}}
\Lambda_j,\\
\label{eq:mjk2}
\mathcal{H}_{\stket{j},\ttket{j'<j}} &=
\frac{1}{\sqrt{j(j+1)j'(j'+1)}}
\Lambda_{j'},\\
\label{eq:mjj}
\mathcal{H}_{\stket{j},\ttket{j}} &=
\frac{1}{j(j+1)}
\left( \sum_{i=1}^{j} \y_i + j^2 \y_{j+1} \right).
\end{align}

We would like to compare the relative sizes of these matrix elements at large $N$.
Using
\begin{align}
\sum_{i=1}^{j} \y_i \to j\braket{\y}
\end{align}
introduces $\mathcal{O}(\Delta_\y)$ errors at small $j$
but becomes exact at $j\gg1$.
At $N\to \infty$, states with
$j,j'\gg1$ are dominant,
for which we can approximate the matrix elements in Eqs.~\ref{eq:m0j}-\ref{eq:mjj}.
Noting that
\begin{align}
\Lambda_j\simeq j( \y -\y_{j+1}) \approx j \mathcal{O}(\Delta_\y),
\end{align}
we obtain
\begin{align}
\label{eq:ma0j}
\mathcal{H}_{\stket{0},\ttket{j}} &=
\mathcal{H}_{\ttket{0},\stket{j}}\approx \mathcal{O}(\Delta_\y)/\sqrt{N},\\
\label{eq:majk1}
\mathcal{H}_{\stket{j},\ttket{j'>j}} &\approx \mathcal{O}(\Delta_\y)/j,\\
\label{eq:majk2}
\mathcal{H}_{\stket{j},\ttket{j'<j}} &\approx \mathcal{O}(\Delta_\y)/j',\\
\label{eq:majj}
\mathcal{H}_{\stket{j},\ttket{j}} &\approx \y.
\end{align}

First, it is clear that the ``off-diagonal'' (${j\neq j'}$) matrix elements in Eqs.~\ref{eq:majk1} and \ref{eq:majk2}
are suppressed by $j,j'$ and so vanish quickly at $j,j'\gg1$
and can be ignored in comparison to the diagonal terms.
The value of the off-diagonal term at the lowest $j,j'=1,2$ is $(\y_1-\y_2)/\sqrt{12}$ from exact expression, Eq.~\ref{eq:mjk1}. Even substituting ${(\y_1-\y_2)\to \Delta_\y}$, we obtain ${0.28 \Delta_\y=0.028\eV}$ at $\Delta_\y=0.1\eV$,
which is still an order of magnitude smaller than the diagonal term in Eq.~\ref{eq:majj} if ${\y\sim0.3\eV}$.
Furthermore, these off-diagonal terms
are $\mathcal{O}(1/\sqrt{N})$ smaller than 
the coupling between the dark and the bright sector, Eq.~\ref{eq:ma0j},
for most of the dark states (large $j$),
so it is a fair approximation to ignore them.
This leads to a great simplification as, treating
Eq.~\ref{eq:ma0j} to be only a weak coupling, it
makes the dark sector a sum of two-level systems (with modified couplings) that can be analytically solved.

Had we considered the Fourier representation for the dark states,
${\ket{\stket{k}} \to \sum_{n=1}^N e^{i 2\pi kn/N} \sig_n^{+} \ket{GS}}$,
${\ket{\ttket{k}} \to \sum_{n=1}^N e^{i 2\pi kn/N} \tau_n^{+} \ket{GS}}$,
we would have obtained 
${\mathcal{H}_{\stket{0},\ttket{j}} = \mathcal{H}_{\stket{l},\ttket{j+l}}}$
with ${j,l \in [1,N-1]}$,
and it would not have been possible to ignore the off-diagonal couplings between the dark states in comparison to the couplings between the bright and the dark sector.

\begin{figure}
\centering
  \includegraphics[width=1\linewidth]{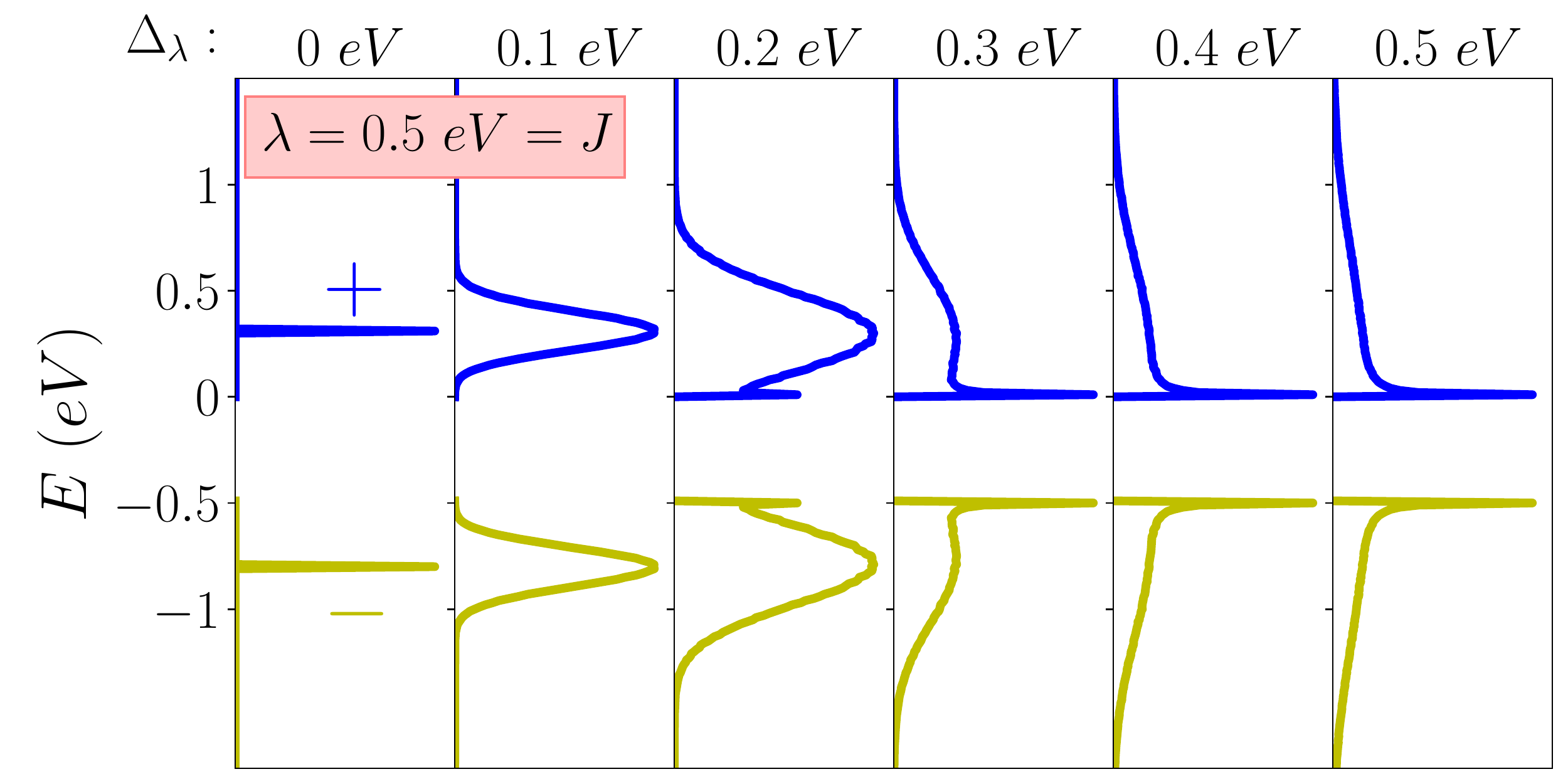}
\caption{ The evolution of the dark bands with the disorder in the spin-orbit couplings $\Delta_\y$.
The density of states of $\pm$ dark bands is shown at $\y=0.5\eV=J$, for a set of $\Delta_\y$ values, $\Delta_\y=0-0.5\eV$ in steps of $0.1\eV$.
}
\label{fig:darkbands}
\end{figure}

\subsection{Dark bands}
\label{sec:}

Assuming the polaritonic states are energetically away from the dark sectors,
the coupling between the dark and the bright states,
Eq.~\ref{eq:ma0j},
will act only as a small perturbation at large $N$. 
For the diagonalization of the dark sectors, 
we can safely ignore it leaving only the diagonal terms (Eq.~\ref{eq:mjj}).
So, the dark bands in the presence of disorder can be analytically calculated, given by,
\begin{align}
\label{eq:darkj}
\ket{\mathcal{D}_{j,\pm}} &= A_{j,\pm} \ket{\stket{j}} \pm B_{j,\pm}\ket{\ttket{j}},\\
E_{j,\pm} &= \frac{1}{2}(-J \pm \sqrt{J^2+4\y_{jj}^2}),\\
A_{j,+} &= \cos(\theta_j), A_{j,-} = \sin(\theta_j),
B_{j,\pm}=A_{j,\mp}, \\
2\theta_j &= \tan^{-1}(2\y_{jj}/J),
\y_{jj}\equiv \mathcal{H}_{\stket{j},\ttket{j}}.
\end{align}

From hereon, we will refer to the two dark bands with their labels $\pm$.
Figure~\ref{fig:darkbands} shows how the distribution of dark band energies $\{E_{j,\pm}\}$, i.e.,
the density of states for the dark bands $\pm$,
evolve with $\Delta_\y$ at $\y=0.5\eV=J$.
At $\Delta_\y=0$, the two bands are flat as expected but 
at a small finite $\Delta_\y$ become Gaussian.
As $\Delta_\y$ increase, the distribution of molecular couplings $\y_{i}$
creates a distribution of dark state couplings $\y_{jj}$ with 
more and more states away from the mean coupling $\y$.
The couplings at and around zero value increase in number
so we obtain the dark bands containing 
 more and more states closer to the bare singlet and triplet dark states.
The same is true for higher couplings that spread the bands away from the corresponding bare state energies, i.e., $E_{j,-}$ are pushed below while $E_{j,+}$ are pushed upwards.
In other words,
a wider distribution of $\y_{i}$ and hence $\y_{jj}$ create 
a wider range of mixing between the bare states, some very close to the pure singlet or triplet and some very close to the maximal $50/50$ mixing.

\subsection{Transitions between dark bands and polaritons}
\label{sec:}

The Hamiltonian of the bright sector is the same as $\h_B$ in Eq.~\ref{eq:hb} under the approximations made above where the coupling between the dark and the bright is assumed weak and ignored in the diagonalization.
$\h_B$
cannot be analytically solved for general parameters, but we can
consider the general form of its eigenstates, given by
\begin{align}
 \label{eq:pol}
\ket{\Phi_x} &= \alpha_x \ket{\stket{0}} + \beta_x \ket{\ttket{0}} + \gamma_x \ket{1_P},
\end{align}
where ${x=LP_\pm,UP}$.

We are interested to find the effects of the perturbation, 
Eq.~\ref{eq:m0j}, on the dynamics of these polaritons and 
the dark states in Eq.~\ref{eq:darkj}. To this end,
the matrix elements of the perturbation between a given pair of states
should give us the probability for the transition between them.
Here, we like to compute ${\mathcal{M}_{j,\pm,x}}$ where,
\begin{align}
\label{eq:mdx}
\mathcal{M}_{j,\pm,x} &=
\braket{\mathcal{D}_{j,\pm}|\hd|\Phi_x},\\
&= \left(\pm B_{j,\pm}\alpha_x +A_{j,\pm}\beta_x\right) 
\mathcal{M}_{j},
\end{align}
where $\mathcal{M}_{j}=\mathcal{H}_{\stket{0},\ttket{j}}=\mathcal{H}_{\ttket{0},\stket{j}}$ is given by
Eq.~\ref{eq:m0j}.

\begin{figure}
\centering
  \includegraphics[width=1\linewidth]{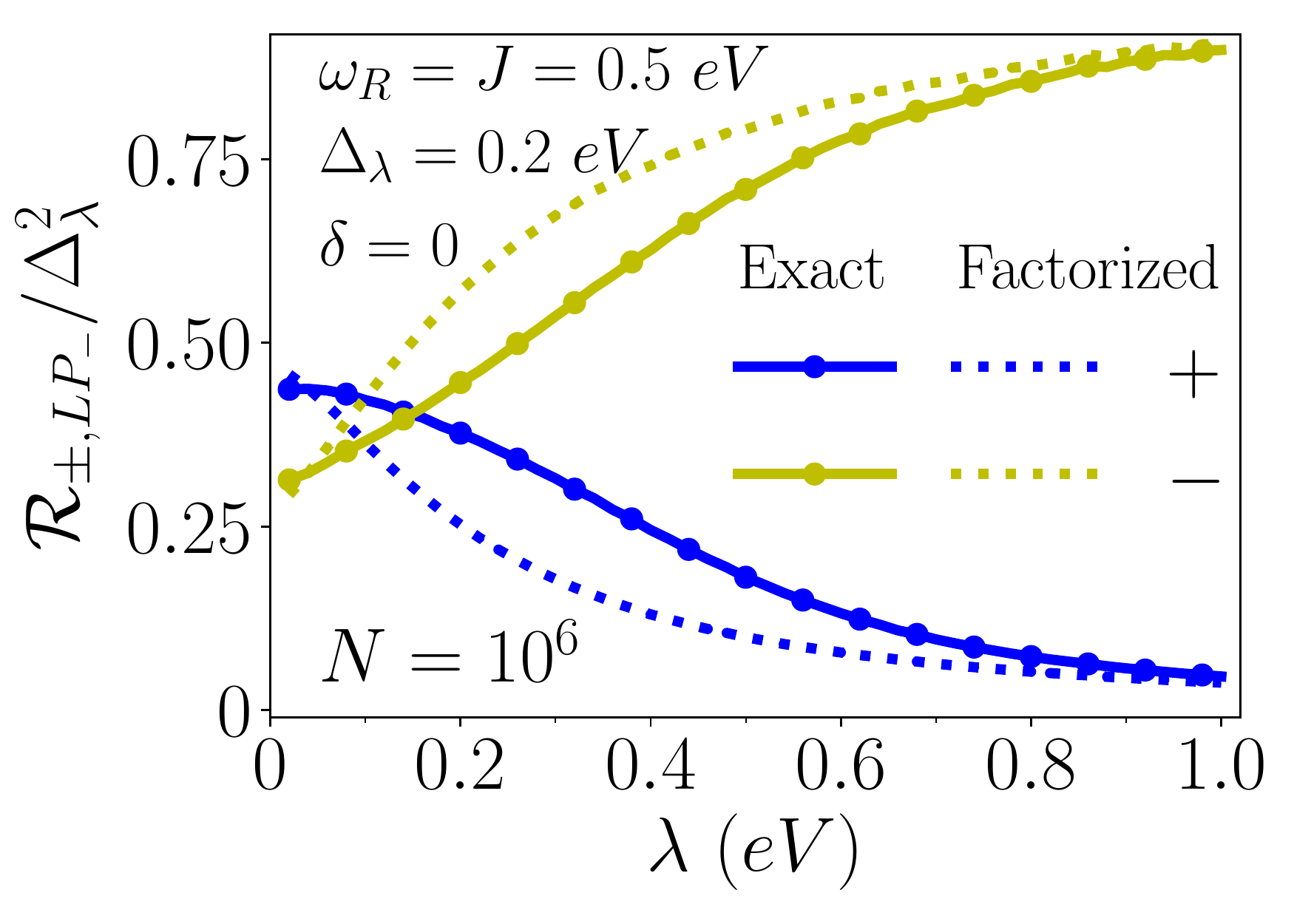}
\caption{
The relaxation of the dark states to $\ket{LP_-}$ induced by the disorder in the spin-orbit coupling.
$\mathcal{R}_{\pm,LP_-}/\Delta_\y^2$ is shown as a function of $\y$
 at $\delta=0$, $\wrr=J=0.5 \eV$.
At $\y=0$, 
the dark states are either singlet or triplet so the interference term vanishes and $\mathcal{R}_{\pm,LP_-}$ are determined only by the weights of the corresponding bright components in $\ket{LP_-}$.
As $\y$ increases, the dark states obtain the missing components and 
the interference between the two transition paths becomes important.
$\mathcal{R}_{+,LP_-}$ ($\mathcal{R}_{-,LP_-}$) decreases (increases) due to 
destructive (constructive) interference.
The factorised case overestimates the interference term.
}
\label{fig:rpmvsl}
\end{figure}

\subsection{Total squared matrix elements $\mathcal{R}_{\pm,x}$}
\label{sec:}

Since 
 the dark state should make a narrow band at typical parameters, 
their energy difference from a given polaritonic state is approximately the same. 
So to calculate the transition rate between the $+$ ($-$) dark band and polariton $\ket{\Phi_x}$,
we can simply sum the corresponding squared matrix elements
over all dark states $\{\ket{\mathcal{D}_{j,+}}\}$ ($\{\ket{\mathcal{D}_{j,-}}\}$).
\begin{align}
\nonumber
\mathcal{R}_{\pm,x}&= \sum_{j=1}^{N-1} |\mathcal{M}_{j,\pm,x}|^2,
\\ \label{eq:rd2p}
 &=\sum_{j=1}^{N-1}
\left(B_{j,\pm}^2\alpha_x^2 + A_{j,\pm}^2\beta_x^2 + \mathcal{I}_{j,\pm,x}\right)
\mathcal{M}_{j}^2, 
\end{align}
where
\begin{align}
\label{eq:interfere}
\mathcal{I}_{j,\pm,x}
&=\pm \alpha_x\beta_x \sin(2\theta_j),
\end{align}
gives the interference between the two transition paths available:
from the dark singlet to the bright triplet and 
from the dark triplet to the bright singlet, in case of a transition from a given dark band ($\pm$) state to a polaritonic state. 
  We will see below that this interference can suppress or enhance the relaxation rate
  quite substantially.

Equation~\ref{eq:rd2p} can be computed numerically, but it can also be approximated as follows.
At ${\y\ll J}$, 
${\mathcal{I}_{j,\pm,x}\sim 0}$
when ${A_{j,+},B_{j,-}\simeq1}$, ${A_{j,-},B_{j,+}\simeq0}$,
and $\ket{\mathcal{D}_{j,+}}\simeq \ket{\stket{j}}$ and $\ket{\mathcal{D}_{j,-}}\simeq\ket{\ttket{j}}$.
Then,
${\mathcal{R}_{+,x}\simeq \beta_x^2  \mathcal{R}}$ and 
${\mathcal{R}_{-,x}\simeq \alpha_x^2 \mathcal{R}}$,
where 
\begin{align}
\label{eq:rs2t}
\mathcal{R} &= \sum_{j=1}^{N-1} |\mathcal{M}_{j}|^2.
\end{align}
From numerical tests at large enough $N$, we find that not only 
$\mathcal{R}$ in Eq.~\ref{eq:rs2t} but also the distribution of the matrix elements $\mathcal{M}_j$
is independent of the mean coupling $\y$, they only depend on the disorder strength $\Delta_\y$. 
This means that,
to see how $\mathcal{R}_{\pm,x}$ behave with $\y$ at a fixed $\Delta_\y$,
 we only need to focus on the distribution of the terms in the prefactor of $\mathcal{M}_j^2$ in Eq.~\ref{eq:rd2p}.

\subsection{Factorization and approximate expressions for $\mathcal{R}_{\pm,x}$}
\label{sec:}

At ${\y\gtrsim J}$, we can approximate $\mathcal{R}_{\pm,x}$
by using the dark bands for the clean system in Eq.~\ref{eq:dark2ls} 
instead of actual dark bands in Eq.~\ref{eq:darkj}. 
We will see that this turns out to be a reasonably good approximation.
So the coefficients and the interference term in Eq.~\ref{eq:rd2p} are assumed to be the same for all states in a given dark band, i.e.,
 ${A_{j,\pm}\to A_{\pm}}$, etc.
It leads to a factorisation,
\begin{align}
\label{eq:rd2pfac}
\mathcal{R}_{\pm,x}
&=\left(B_{\pm}^2\alpha_x^2 + A_{\pm}^2\beta_x^2 + \mathcal{I}_{\pm,x}\right)\mathcal{R}.
\end{align}
Thus, the problem reduces to the computation of the total squared matrix element $\mathcal{R}$
between the bright singlet and the whole dark triplet space (or vice versa) along with the diagonalization of a two-level dark and three-level polariton system
to compute the prefactor in Eq.~\ref{eq:rd2pfac}.


\subsection{Numerical Results}
\label{sec:}

Figure~\ref{fig:rpmvsl} presents the numerical results for the total squared matrix element between the dark bands $\pm$ and the lowest polaritonic state $\ket{LP_-}$.
$\mathcal{R}_{\pm,LP_-}$ from Eq.~\ref{eq:rd2p} for $x=LP_-$ is 
shown
as a function of $\y$ at ${\wrr=J=0.5\eV}$, ${\delta=0,\Delta_\y=0.2\eV}$, and $N=10^6$.
For comparison, the factorized version in Eq.~\ref{eq:rd2pfac} is also shown.
The results are well converged with respect to the system size $N$.
We note that the factorized results are reasonably good approximation.
To understand the exact results,
we will first look at the behavior of the factorized version and then see why the exact results differ.

We will see below that the factor $\mathcal{R}$ in Eq.~\ref{eq:rd2pfac} is essentially independent of $\y$
so the behavior of the factoried case as a function of $\y$
can be understood by looking at how the coefficients in the prefactor of $\mathcal{R}$ in Eq.~\ref{eq:rd2pfac},  
in particular, the interference term $\mathcal{I}_{\pm,x}$,
change with $\y$.
For $x=LP_-$, the coefficients $\alpha_x,\beta_x$ have different signs (at $\y,\wrr\geq0$),
so the interference term can be written as
\begin{align}
\label{eq:ainterfere}
\nonumber
\mathcal{I}_{\pm,x}
&=\mp |\alpha_x\beta_x| \sin(2\theta),\\ \nonumber
&\propto \mp \sin(2\theta) = \mp \y/\sqrt{\y^2 + (J/2)^2},
\end{align}
which means that the two paths will interfere constructively for the $-$ and destructively for the $+$ dark band.

As discussed before for the $\y\ll J$ limit,
at small $\y\sim0$ in Fig.~\ref{fig:rpmvsl}, either component significantly dominates in the two dark bands
(the singlet dominates in the $+$ band and the triplet dominates in the $-$ band),
reducing the size of $\mathcal{I}_{\pm,x}$.
So $\mathcal{R}_{\pm,LP_-}$ are determined primarily from the size of the 
bright exciton components in $\ket{LP_-}$.
At the parameters considered in Fig.~\ref{fig:rpmvsl},
the bright triplet component of $\ket{LP_-}$ is slightly larger than its bright singlet component,
so the $+$ band with dark singlet character relaxes more quickly.
That is,
 $\mathcal{R}_{+,LP_-}$
is slightly larger than $\mathcal{R}_{-,LP_-}$ at $\y\sim0$.
However, as $\y$ increases, the two bright exciton components in $\ket{LP_-}$ increase slightly 
but the $\pm$ dark bands acquire more and more mixed character, 
so the interference term quickly becomes more and more important,
 explaining the increase and decrease in $\mathcal{R}_{\pm,LP_-}$.

Compared to the exact results, 
the factorized version overestimates the interference effect.
The reason is simple.
A distribution in the couplings $\{\y_i\}$ creates a similar 
distribution for the matrix elements
$\y_{jj}=\mathcal{H}_{\stket{j},\ttket{j}}$ 
(see Fig.~\ref{fig:jdistrib} in the tAppendix~\ref{ssec:suppress}),
but it in turn creates a strongly distorted distribution 
for ${\sin(2\theta_j)=\y_{jj}/\sqrt{\y_{jj}^2+(J/2)^2}}$ 
(that determines the interference term $\mathcal{I}_{j,\pm,x}$)
that has lower average compared to the factorized model [${\sin(2\theta)=\y/\sqrt{\y^2+(J/2)^2}}$] at $\y>0$, with the difference 
first increasing (at $\y\lesssim J/2$) and then decreasing (at $\y\gtrsim J/2$) with an increase in $\y$.
This is further discussed in the Appendix~\ref{ssec:suppress} where the distributions are also shown.

\begin{figure}
\centering
  \includegraphics[width=1\linewidth]{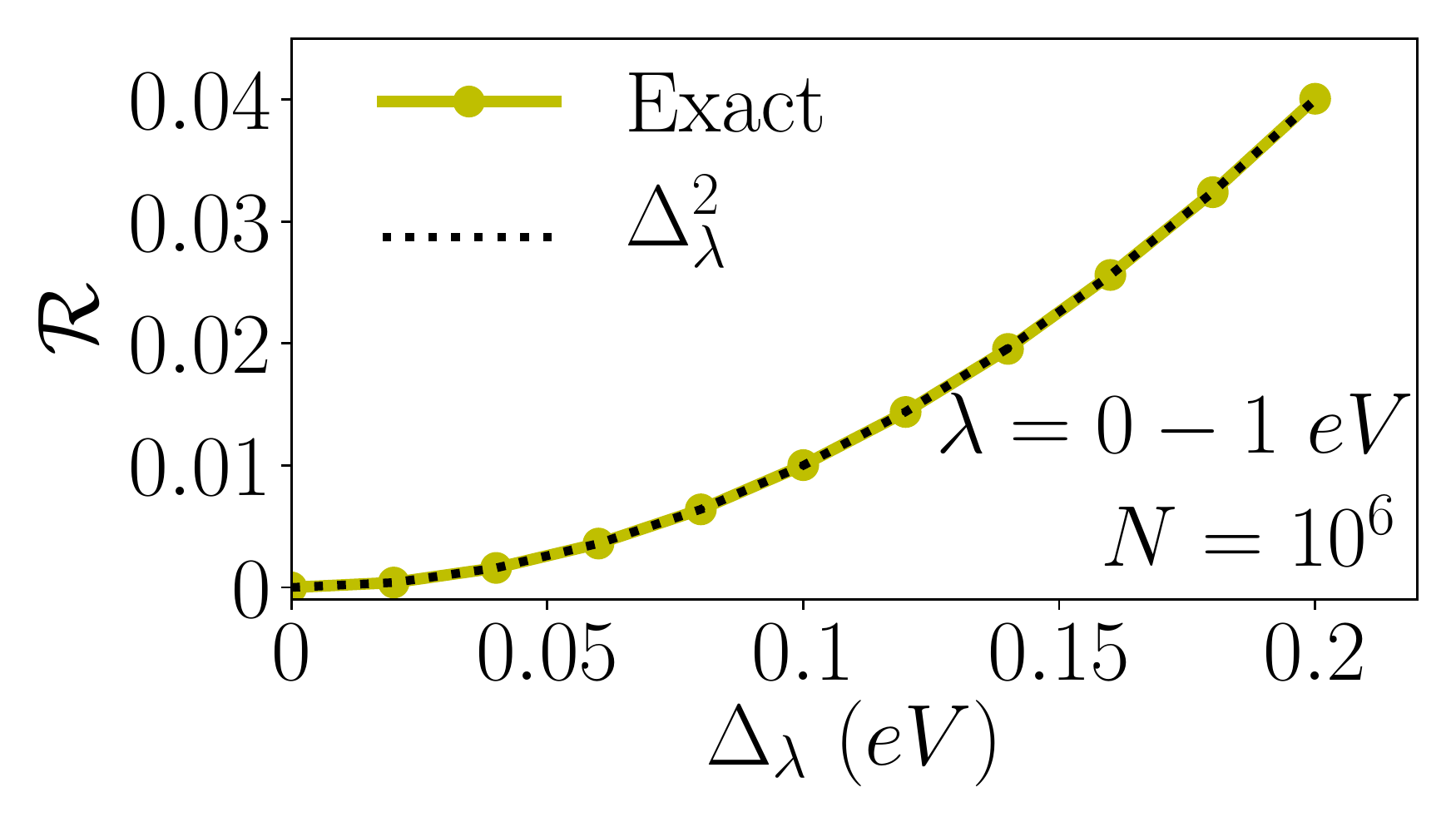}
\caption{
The squared matrix elements $\mathcal{R}$ 
that determines the transition rates 
between the dark states and polaritons
as a function of 
the disorder in the spin-orbit coupling $\Delta_\y$.
Analytical results $\mathcal{R}\simeq\Delta_\lambda^2$ also shown for comparison.
}
\label{fig:rvsdl}
\end{figure}

\subsection{Analytical results for $\mathcal{R}$ at large $N$}
\label{sec:}

Let us now consider the factor $\mathcal{R}$ in Eq.~\ref{eq:rd2pfac}.
At ${N\to \infty}$,
an approximate but quite accurate expression for $\mathcal{R}$ can be obtained. 
Replacing
\begin{align}
\sum_{i=1}^{j} \y_i \to j\braket{\y},
\end{align}
in Eq.~\ref{eq:rs2t} [with ${\mathcal{M}_j=\mathcal{H}_{\stket{0},\ttket{j}}}$
in Eq.~\ref{eq:m0j}],
where $\braket{\y}$ is the mean value of the spin-orbit coupling,
introduces $\mathcal{O}(\Delta_\y)$ errors at small $j$
but becomes exact at $j\gg1$. 
With this approximation, we obtain
\begin{align}
\label{eq:r2}
\mathcal{R}
&= \frac{1}{N} \sum_{j=1}^{N-1} \frac{j}{(j+1)}
\left(
\braket{\y}^2 +  \y_{j+1}^2 - 2 \braket{\y} \y_{j+1}
\right).
\end{align}

Again, replacing ${j/(j+1)\to 1}$ affects only a few terms with small $j\sim 1$,
and should not affect the results in the thermodynamic limit, $N\to\infty$.
Making this substitution in Eq.~\ref{eq:r2}
and noting that
\begin{align}
\sum_{j=1}^{N-1} \y_{j+1} &=(N-1)\braket{\y},\\
\sum_{j=1}^{N-1} \y_{j+1}^2&=(N-1)\braket{\y^2},
\end{align}
we obtain,
\begin{align}
 \nonumber
\mathcal{R}
& \simeq 
\frac{N-1}{N}
\left(
\braket{\y}^2+ \braket{\y^2} - 2\braket{\y}^2
\right),\\ \nonumber
& \simeq 
\left(
\braket{\y^2} - \braket{\y}^2
\right),\\  \label{eq:dl2}
& = \Delta_\y^2.
\end{align}

\subsection{Comparison of analytical and numerical results for $\mathcal{R}$}
\label{sec:}

Figure~\ref{fig:rvsdl}
shows the exact results for $\mathcal{R}$ at $N=10^6$,
numerically computed from Eqs.~\ref{eq:rs2t} and \ref{eq:m0j},
as well as the analytical expression, Eq.~\ref{eq:dl2}.
The difference is invisible.
The results also do not depend on $\y$, checked for ${\y=0-1\eV}$.
It is worth reminding that the results do not depend on the explicit values of other parameters $\delta, \wrr$, etc., as
$\mathcal{R}$ is \emph{not} the squared matrix element for the eigenstates but for the bright and dark states of different exciton types.

\subsection{A simplified formula for $\mathcal{R}_{\pm,x}$}
\label{sec:}

Using Eqs.~\ref{eq:dl2} and \ref{eq:rd2pfac},
we obtain a very simple expression,
\begin{align}
\label{eq:rad2p}
\mathcal{R}_{\pm,x}
&\simeq \left(B_{\pm}\alpha_x \pm A_{\pm}\beta_x\right)^2  \Delta_\y^2,
\end{align}
which can now be computed from the two dark (Eq.~\ref{eq:dark2ls}) and three polaritonic eigenstates, 
eliminating all complications.
For the polaritonic states ${x=LP_{\pm}}$, we can even use the expressions obtained in sec.~\ref{sec:approx},
i.e., Eq.~\ref{eq:analv}, which makes Eq.~\ref{eq:rad2p} an analytical expression.
It can now be used with an appropriate open quantum system method to calculate, for instance, the population dynamics of the dark states and polaritons.


\section{Summary and outlook}
We investigate the role of the spin-orbit coupling between the singlet and the triplet molecular excitons in organic microcavities. 
We decouple the system into a bright sector that forms polaritons and a dark sector containing the dark exciton states that are decoupled from the cavity and form two dark bands.
We show that the spin-orbit coupling splits the lower polariton
 into two branches that can be observed in the optical absorption spectrum. 
 We find that when the cavity is in resonance with the triplet state, we obtain \emph{triplet} polaritons.
 We also see the effects of disorder in energies and couplings on the absorption spectrum and see that the experiments should be able to find the spin-orbit induced splitting at typical disorder strengths.
The spin-orbit coupling couples the bright singlet to the bright triplet and 
the dark singlet to the dark triplet sectors
but does not couple the bright to the dark sectors.
We show that the disorder in the spin-orbit coupling does exactly that
and hence can be potentially very useful in optoelectronic devices where
dark population is desired to be transferred to the polaritons.
Considering it to be a weak coupling,
we calculate the squared transition matrix elements between the dark bands and polaritons and also find approximate analytical expressions for them that can be employed to
calculate the population dynamics.

There are plenty of organic molecules that 
have been proven to exhibit large enough spin-orbit coupling
in electroluminescent devices 
where
an efficient ISC along with
a large radiative phosphorescent rate,
or a large RISC rate in TADF,
is observed.
Some examples include heavy metal complexes~\cite{yersin2011}
containing Ir, Pt, Os, Cu, Ag etc.,
such as
PtOEP (platinum octaethylporphyrin)~\cite{baldo1998},
Ir(ppy)$_3$, Btp$_2$Ir(acac), FIrpic ~\cite{kawamura2005},
Pt(4,6-dFppy)(acac) and Ir(4,6-dFppy)$_2$(acac)~\cite{Rausch2010},
Ir(dm-2-piq)$_2$(acac), Cu(POP)(pz$_2$BH$_2$)~\cite{yersin2011},
and mixed-ligand iradium complexes
Ir(Phbib)(ppy)Cl, Ir(Mebib)(mppy)Cl, Ir(Mebib)(ppy)Cl, etc.~\cite{ObaraIC06}.
Since these molecules are excellent at photoemission,
they also have a large dipole matrix element 
required to achieve strong matter-light coupling,
and thus 
are promising candidates to
demonstrate
the lower polariton splitting 
or an enhanced dark state to polariton relaxation 
as predicted in this work.

Besides the experimental scrutiny of our results, 
various unexplored avenues for future theoretical work also exist
as described in the following.  
We know that
the triplet exciton has a net spin and hence a magnetic moment, 
so
$\ket{LP_\pm}$ would borrow it (proportional to the triplet weight)
and should thus be manipulated using an external magnetic field.
It would be interesting to consider all three triplet states in the model 
and explore the response of the system to such a field.
When both the dark bands are able to relax to $\ket{LP_-}$,
it can bring down the threshold pump power for creating their condensate or lasing.
Considering it an open quantum system with an appropriate bath spectral density to study how the spin-orbit coupling and its disorder affect this threshold is intriguing. 
Due to the polaron decoupling at large $N$,
we ignored the vibrations and vibronic coupling in our model.
Exploring the interplay between vibrations and spin-orbit coupling at small $N$
can also be a future work.

 \com{A disorder in the spin-orbit coupling couples the bright and the dark sectors weakly. We calculate the squared transition matrix elements between the dark bands and polaritons and also find approximate analytical expressions for them that can be employed to
calculate the population dynamics. 
}

\begin{acknowledgments}
MAZ thanks Rukhshanda Naheed for fruitful discussions.
\end{acknowledgments}

\appendix*

\com{
\section{Triplet polariton composition}
\label{ssec:tp}
The Hoppfield coefficients of the photon $\ket{1_P}$ and the bright triplet state 
$\ket{\ttket{0}}$
obtained from Eq.~\ref{eq:trippol} 
are shown in Fig.~\ref{fig:tp} as a function of $\y$ at $\wrr=0.5\eV$.
We see that, at $\y<\wrr$, the $\ket{\ttket{0}}$ dominates while the state becomes more photon like at $\y>\wrr$ where $\ket{1_P}$ component becomes larger.
The maximal mixing occurs at $\y=\wrr$.
\begin{figure}
\centering
   \includegraphics[width=0.8\linewidth]{triplet-polaritons}
\caption{The weights of photon and triplet exciton components in triplet polariton $\ket{LP_+}_T$ as a function of $\y$ at $\wrr=0.5\eV$ (and any J with $\delta=-J$). 
}
\label{fig:tp}
\end{figure}
}

\section{SUPPRESSION OF THE INTERFERENCE BETWEEN THE TWO TRANSITION PATHS DUE TO DISORDER}
 \label{ssec:suppress}


\begin{figure}
\centering
  \includegraphics[width=1\linewidth]{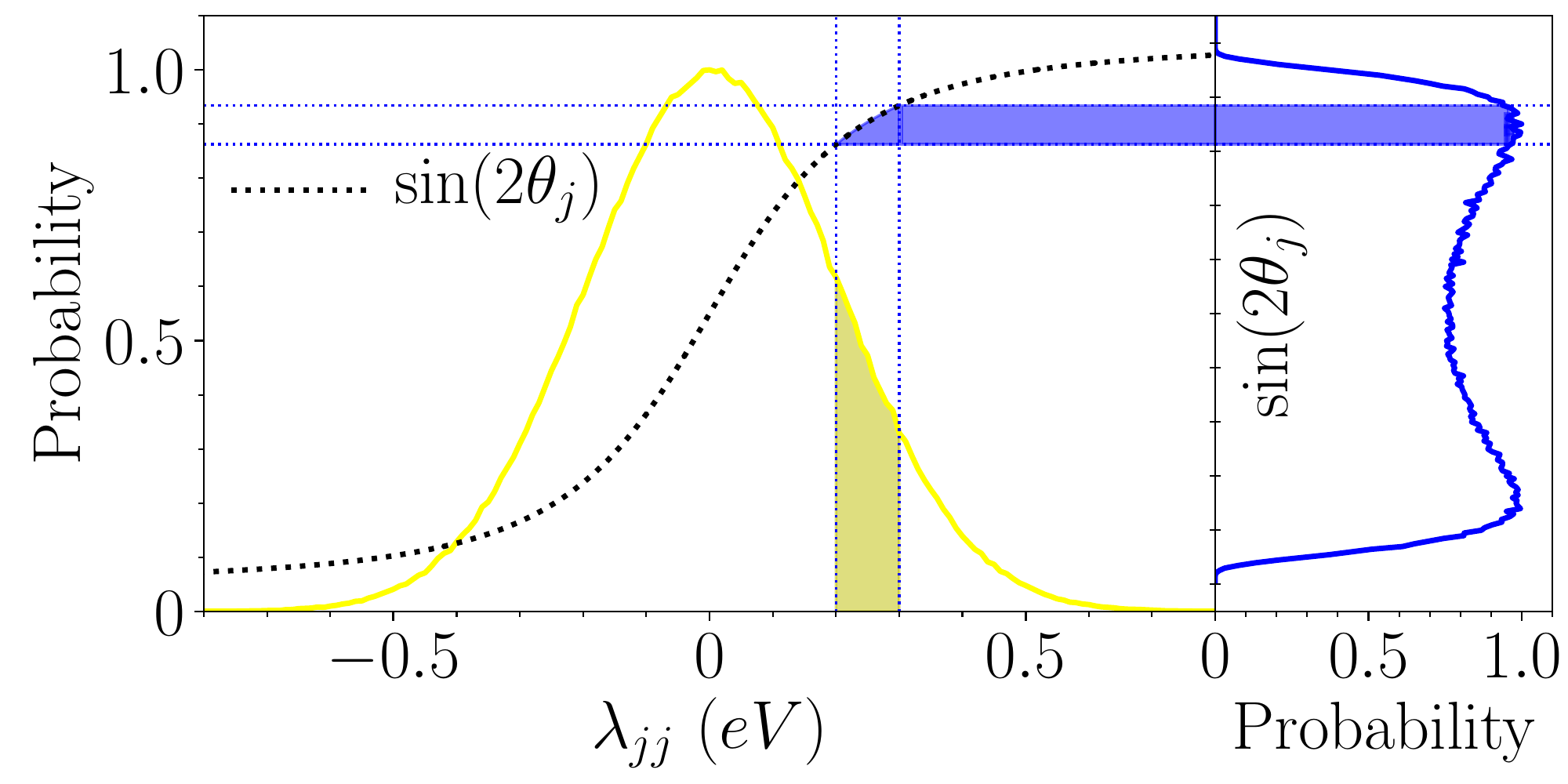}
\caption{(Color online) An illustration of how a distribution of $\y_{jj}$ is mapped onto a distribution of $\sin(2\theta_j)$.
The number of entries in the $\y_{jj}$ in the yellow (gray) shaded region go to the blue (black) slot in the 
$\sin(2\theta_j)$ distribution.
The entries on the right of the yellow (gray) shaded region in the $\y_{jj}$ distribution shown or in any other distribution that we consider (where $\y>0$) all go to the narrow region above the blue (black) shaded region below the thick black dotted curve. 
}
\label{fig:jtransform}
\end{figure}

\begin{figure}
\centering
  \includegraphics[width=1\linewidth]{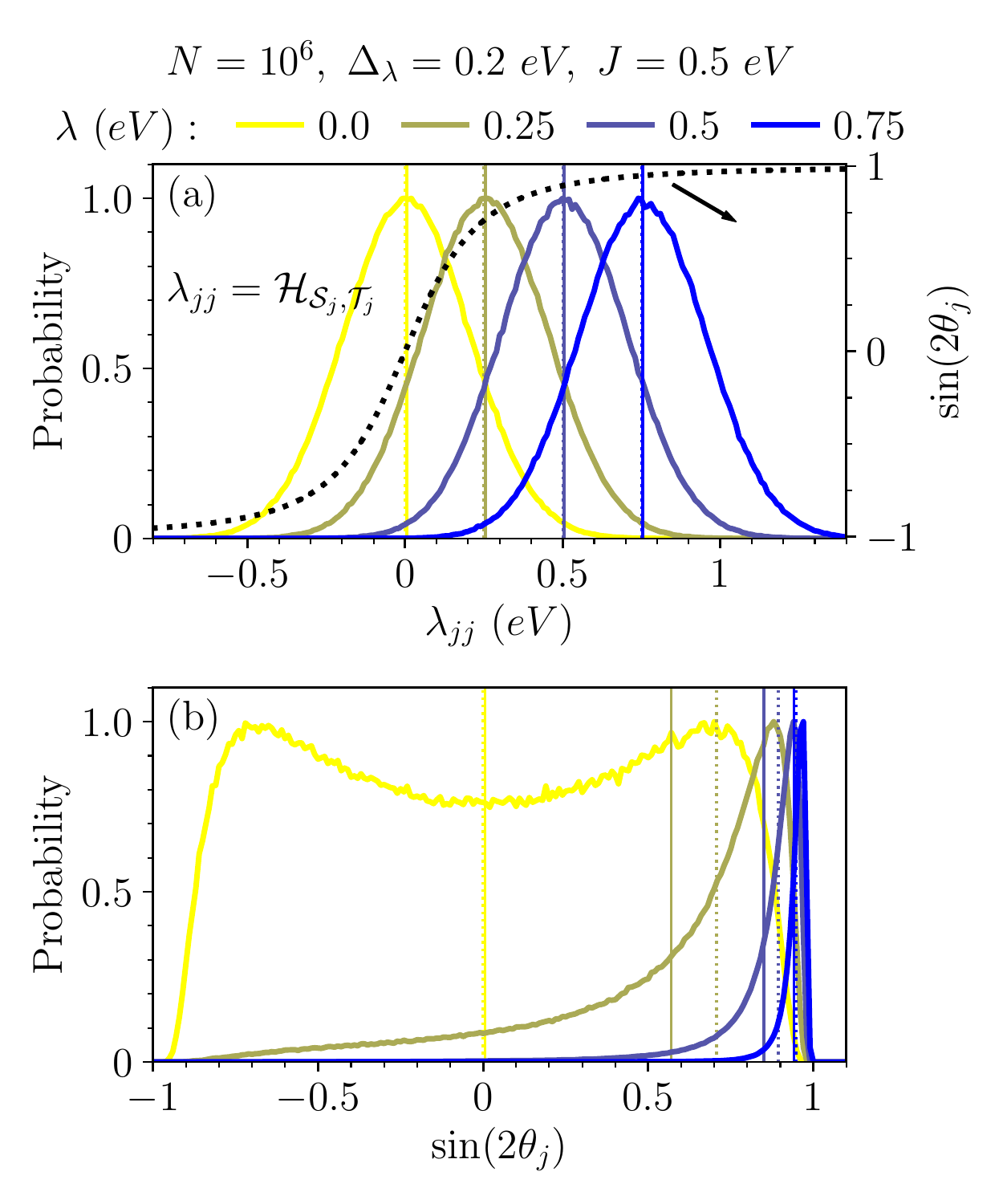}
\caption{(Color online)
The
 influence of the distribution of couplings $\{\y_i\}$ on 
 the couplings between the dark states $\y_{jj}$
 and the interference terms $\sin(2\theta_j)$.
 The distributions are calculated at $N=10^6$, $\Delta_\y=0.2\eV$ and $J=0.5\eV$
for $\y=0,0.25,0.5,0.75\eV$ with line colour changing from yellow to blue (gray to black). 
The distributions are normalised by their peak value for easier comparison.
Solid and dotted vertical lines show the mean value of the distribution and the value for the factorised case, Eq.~\ref{eq:rd2pfac}.
(a)
The distributions of the matrix elements between the singlet and the triplet dark states
$\y_{jj}$.
The distributions do not seem to differ much from the original gaussians for  $\{\y_i\}$ (not shown here).
The thick dotted curve shows the dependence of the interference term ${\sin(2\theta_j)}$ on $\y_{jj}$---the values are shown on the vertical axis on the right side as indicated by a small arrow. 
(b) The distributions of $\sin(2\theta_{jj})$ corresponding to the distributions of $\y_{jj}$ shown in (a).
The mean value for the distributions (solid vertical lines) is lower from that of the factorised case (dotted vertical lines) at $\y\geq 0.25\eV$ but the difference decreases with an increase (or decrease, see text) in $\y$. 
}
\label{fig:jdistrib}
\end{figure}

First, let us understand how a distribution of couplings $\y_{jj}$ transforms into a distribution of the interference terms, $\sin(2\theta_j)$.
Figure~\ref{fig:jtransform} shows a $\y_{jj}$ distribution that is centered around zero
along with 
the function $\sin(2\theta_j)$.
The number of elements in the $\y_{jj}$ distribution in a window $0.2\eV>\y_{jj}>0.3\eV$
[yellow (gray) shaded region] will move to the blue (black) shaded region on the $\sin(2\theta_j)$ distribution that is shown on the right. 
The elements at $\y_{jj}>0.5\eV$ are very few in this distribution but can be half of the total for a distribution with $\y\sim0.5\eV$. 
Following the above rule, 
they all move to a very narrow region close to the extreme value of $\sin(2\theta_j)=1$ making a sharp peaked distribution there.
The entries close to ${\y_{jj}=0}$ will spread over a larger region on the ${\sin(2\theta_j)}$ distribution because it has a sharp change around $\y_{jj}=0$, 
dispersing the distribution shown here (with mean at ${\y=0}$) symmetrically on either side but
creating a 
long tail for slightly shifted distributions.
This is seen in Fig.~\ref{fig:jdistrib},
where
the distributions of $\y_{jj}$ and ${\sin(2\theta_j)}$
are shown
at ${N=10^6}$, ${\Delta_\y=0.2\eV}$, ${J=0.5\eV}$
and a set of $\y$ values, ${\y=0,0.25,0.5, and 0.75\eV}$.
The mean value of the distribution and the value for the factorized case (that is equivalent to assuming no disorder in $\y_i$), Eq.~\ref{eq:rd2pfac},
are also shown by solid and dotted vertical lines.

As explained using Fig.~\ref{fig:jtransform},
compared to the input distributions $\y_{jj}$,
the distributions of ${\sin(2\theta_j)}$ are 
spread more near zero but
shifted more away from zero, resulting in flattened and broadened distribution at $\y=0$ (red), 
but creating a long tail at $\y=0.25\eV$ and pushing the whole distributions at $\y=0.5 and 0.75\eV$ close to $1$ 
on the ${\sin(2\theta_j)}$ scale.
The asymmetry introduced lowers
the mean value for the distributions (solid vertical lines) 
from the factorized case (dotted vertical lines) for all $\y>0$ cases shown,
with the difference decreasing at higher $\y$.

We can now understand that 
at $0<\y<J/2=0.25\eV$ (distributions not shown), the difference keeps increasing with an increase in $\y$.
This increase and decrease of the difference between the exact and the factorize case is reflected in Fig.~\ref{fig:rpmvsl},
and shows that the interference is, 
on average, suppressed by the disorder in the couplings and the suppression is maximum around $\y=J/2$.
The reason that the factorized model overestimates the interference effect in Fig.~\ref{fig:rpmvsl} is that it ignores this suppression.

\label{Bibliography}
\bibliographystyle{apsrev4-1}
 
%
  \end{document}